\documentclass[acmsmall, nonacm, 10pt]{acmart}

\AtBeginDocument{%
  }

\copyrightyear{}
\acmYear{}
\acmDOI{}
\usepackage{ifthen}
\usepackage{xspace}
\usepackage{subcaption}
\usepackage[normalem]{ulem}
\usepackage[nolist,nohyperlinks]{acronym}
\usepackage{color}
\usepackage{listings}
\usepackage[utf8]{inputenc}

\graphicspath{{./images/}}

\newcommand{\etal}{et\,al.\xspace}

\newboolean{authorcomments}
\setboolean{authorcomments}{true}
\newcommand{\bssm}[1]{\ifthenelse{\boolean{authorcomments}}{{\color{blue}\textit{[#1 -ssm-]}}\xspace}{}}
\newcommand{\rssm}[1]{\ifthenelse{\boolean{authorcomments}}{{\color{red}\textit{[#1 -ssm-]}}\xspace}{}}
\newcommand{\gssm}[1]{\ifthenelse{\boolean{authorcomments}}{{\color[rgb]{0,0.6,0}\textit{[#1 -ssm-]}}\xspace}{}}

\newboolean{long}

\newcommand{\lvsv}[2]{\ifthenelse{\boolean{long}}{#1}{#2}}
\newboolean{anonymous}
\setboolean{anonymous}{true}
\newcommand{\onlyan}[1]{\ifthenelse{\boolean{anonymous}}{#1}{}} 
\newcommand{\onlynonan}[1]{\ifthenelse{\boolean{anonymous}}{}{#1}} 
\newcommand{\annonan}[2]{\ifthenelse{\boolean{anonymous}}{#1}{#2}} 

\newcommand{\chapterquote}[2]{
	\begin{flushright}
		\begin{minipage}{0.68\linewidth}
			\begin{flushright}
				\textit{#1}\\\smaller--- #2
			\end{flushright}
		\end{minipage}
	\end{flushright}
	\vspace{4.5ex}
}

\newcommand{\cinco}{\textsc{Cin\-co}\xspace}
\newcommand{\dime}{\textsc{Dime}\xspace}

\definecolor{shadecolor}{rgb}{0.9,0.9,0.9}

\newcommand{\webstory}{\emph{WebStory~}}

\begin{acronym}
    \acro{Kuerzel}[Kurzform]{Langform}
    \acro{add}[ADD]{Algebraic Decision Diagram}
    \acro{alex}[ALEX]{Automata Learning Experience}
    \acro{api}[API]{Application Programming Interface} \acused{api}
    \acro{ast}[AST]{abstract syntax tree}
    \acro{aws}[AWS]{Amazon Web Services} \acused{aws}
    \acro{bpmn}[BPMN]{Business Process Modeling Notation}
    \acro{ci/cd}[CI/CD]{Continuous Integration / Deployment} \acused{ci/cd}
    \acro{ci}[CI]{Continuous Integration}
    \acro{cinco}[\cinco{}]{\cinco{}} \acused{cinco}
    \acro{com}[com.]{commercial}
    \acro{cpd}[CPD]{\cinco{} Product Definition}
    \acro{cpgp}[CPGP]{\cinco{} Product Generation Process}
    \acro{crud}[CRUD]{Create Read Update Delete} \acused{crud}
    \acro{css}[CSS]{Cascading Style Sheets}
    \acro{ct-api}[CT-API]{\cinco{} Transformation API}
    \acro{ctl}[CTL]{Comutation Tree Logic} \acused{ctl}
    \acro{cbyc}[CbyC]{correctness by construction}
    \acro{cobots}[Cobots]{Collaborative Robots}
    \acro{css}[CSS]{Cascade Style Sheet} \acused{css}
    \acro{cudd}[CUDD]{Colorado University Decision Diagram Package}
    \acro{dad}[DAD]{\dime{} Application Descriptor} 
    \acro{dag}[DAG]{directed acyclic graph} \acused{dag}
    \acro{ddd}[DDD]{Domain-Driven Design}
    \acro{dsl}[DSL]{Domain-specific Language}
    \acro{edsl}[eDSL]{embedded Domain-specific Language} \acused{edsl}
    \acro{emf}[EMF]{Eclipse Modeling Framework}
    \acro{eml}[EML]{enterprise modeling language}
    \acro{eol}[EOL]{Epsilon Object Language}
    \acro{eti}[ETI]{Electronic Tool Integration Platform}
    \acro{ffi}[FFI]{Foreign-Function Interface} \acused{ffi}
    \acro{gro}[GRO]{General Register Office}
    \acro{gui}[GUI]{graphical user interface} \acused{gui}
    \acro{hmi}[HMI]{Human Machine Interface}
    \acro{html}[HTML]{Hypertext Markup Language}
    \acro{ide}[IDE]{Integrated Development Environment} \acused{ide}
    \acro{ime}[IME]{Integrated Modeling Environment}
    \acro{ipc}[IPC]{Industrial PC}
    \acro{ipsum}[IPSUM]{Industrial Process Software Utilizing Modelization} \acused{ipsum}
    \acro{iac}[IaC]{Infrastructure as Code}
    \acro{iot}[IoT]{Internet of Things} \acused{iot}
    \acro{iso}[ISO]{International Organisation for Standardization} \acused{iso}
    \acro{kts}[KTS]{Kripke Transition System}
    \acro{lcd}[LCD]{Low-Code Development}
    \acro{lde}[LDE]{Language-Driven Engineering}
    \acro{lop}[LOP]{Language-Oriented Programming}
    \acro{mde}[MDE]{Model-Driven Engineering} \acused{mde}
    \acro{mgl}[MGL]{Meta Graph Language}
    \acro{ml}[ML]{} \acused{ml}
    \acro{mps}[MPS]{Meta Programming System}
    \acro{msl}[MSL]{Meta Style Language}
    \acro{mvc}[MVC]{model, view, control}
    \acro{ocr}[OCR]{Optical Character Recognition}
    \acro{ocs}[OCS]{Online Conference Service}
    \acro{os}[OS]{open-source}
    \acro{ota}[OTA]{\emph{One Thing Approach}}
    \acro{plc}[PLC]{Programmable Logic Controller}
    \acro{pid}[P\&ID]{Piping and Instrumentation Diagram}
    \acro{psl}[PSL]{Purpose-Specific Language}
    \acro{rcp}[RCP]{Rich Client Platform}
    \acro{rpc}[RPC]{Remote Procedure Call}
    \acro{sde}[SDE]{Service Definition Environment}
    \acro{sib}[SIB]{Service-Independent Building Block}
    \acro{slg}[SLG]{Service Logic Graph}
    \acro{sos}[SOS]{Structural Operational Semantics}
    \acro{sql}[SQL]{Structured Query Language} \acused{sql}
    \acro{spa}[SPA]{single page application}
    \acro{sul}[SUL]{system under learning}
    \acro{tcp/ip}[TCP/IP]{Transmission Control Protocol/Internet Protocol}
    \acro{ui}[UI]{user interface} \acused{ui}
    \acro{ux}[UX]{user experience} \acused{ux}
    \acro{uml}[UML]{Unified Modeling Language}
    \acro{url}[URL]{Uniform Resource Locator} \acused{url}
    \acro{vhdl}[VHDL]{Very High Speed Integrated Circuit Hardware Description Language}
    \acro{xmdd}[XMDD]{eXtreme model-driven design}
    \acro{xml}[XML]{Extensible Markup Language}
    \acro{yaml}[YAML]{YAML Ain't Markup Language} \acused{yaml}
    \acro{mide}[mIDE]{Mindset-Supporting Integrated Development Environment}
    \acro{jabc}[jABC]{Java Application Building Center}
\end{acronym}

\begin{document}

\title[Language-Driven Engineering]{Language-Driven Engineering}
\subtitle{An Interdisciplinary Software Development Paradigm}

\renewcommand*{\thefootnote}{\arabic{footnote}}
\author{Bernhard Steffen}
\email{bernhard.steffen@cs.tu-dortmund.de}
\orcid{0000-0001-9619-1558}
\authornote{TU Dortmund University}

\author{Tiziana Margaria}
\email{}
\orcid{0000-0002-5547-9739}
\authornote{University of Limerick}

\author{Alexander Bainczyk}
\email{}
\orcid{0009-0005-2858-2174}
\authornotemark[1]

\author{Steve Bo{\ss}\-elmann}
\email{}
\orcid{0000-0001-9454-6757}
\authornotemark[1]

\author{Daniel Busch}
\email{}
\orcid{0009-0000-1832-9857}
\authornotemark[1]

\author{Marc Driessen}
\email{}
\orcid{0000-0002-2284-4808}
\authornote{Springer Nature BV}

\author{Markus Froh\-me}
\email{}
\orcid{0000-0001-6520-2410}
\authornotemark[1]

\author{Falk Howar}
\email{}
\orcid{0000-0002-9524-4459}
\authornotemark[1]

\author{Sven J\"orges}
\email{}
\orcid{0000-0002-3071-7163}
\authornote{Dortmund University of Applied Sciences and Arts}

\author{Marvin Krause}
\email{}
\authornotemark[1]

\author{Marco Krumrey}
\email{}
\authornotemark[1]

\author{Anna-Lena Lamprecht}
\email{}
\orcid{0000-0003-1953-5606}
\authornote{University of Potsdam}

\author{Michael Lybecait}
\email{}
\authornotemark[1]

\author{Alnis Murtovi}
\email{}
\authornotemark[1]

\author{Stefan Naujokat}
\email{}
\orcid{0000-0002-6265-6641}
\authornotemark[1]

\author{Johannes Neubauer}
\email{}
\authornotemark[1]

\author{Alexander Schieweck}
\email{}
\orcid{0000-0002-5008-9168}
\authornotemark[2]

\author{Jonas Sch\"urmann}
\email{}
\orcid{0000-0002-1587-0549}
\authornotemark[1]

\author{Steven Smyth}
\email{steven.smyth@tu-dortmund.de}
\orcid{0000-0003-2470-0880}
\authornotemark[1]

\author{Barbara Steffen}
\email{}
\orcid{0000-0002-0825-8490}
\authornotemark[1]

\author{Fabian Storek}
\email{}
\authornotemark[1]

\author{Tim Tegeler}
\email{}
\orcid{0000-0002-8271-9072}
\authornotemark[1]

\author{Sebastian Teum\-ert}
\email{}
\orcid{0000-0002-6483-3162}
\authornotemark[1]

\author{Dominic Wirkner}
\email{}
\orcid{0009-0004-2842-0311}
\authornotemark[1]

\author{Philip Zweihoff}
\email{}
\orcid{0000-0001-5851-8521}
\authornotemark[1]

\renewcommand{\shortauthors}{Steffen et al.}

\makeatletter
\let\@authorsaddresses\@empty
\makeatother

\renewcommand\footnotetextcopyrightpermission[1]{}

\begin{abstract}
We illustrate how purpose-specific, graphical modeling enables application experts with different levels of expertise to collaboratively design and then produce complex applications using their individual, purpose-specific modeling language. 
Our illustration includes seven graphical Integrated Modeling Environments (IMEs) that support full code generation, as well as four browser-based applications that were modeled and then fully automatically generated and produced using \dime{}, our most complex graphical IME. 
While the seven IMEs were chosen to illustrate the types of languages we support with our Language-Driven Engineering (LDE) approach, the four \dime{} products were chosen to give an impression of the power of our LDE-generated IMEs. 
In fact, Equinocs, Springer Nature's future editorial system for proceedings, is also being fully automatically generated and then deployed at their Dordrecht site using a deployment pipeline generated with Rig, one of the IMEs presented. 
Our technology is open source and the products presented are currently in use.
\end{abstract}


\fancyfoot[RO,LE]{}
\renewcommand{\thefootnote}{\arabic{footnote}}
\maketitle
\fancyfoot[RO,LE]{}


\section{Introduction}
\label{sec:introduction}

\chapterquote{The market for automobiles will never grow beyond one million cars, for a very simple reason: Who would
	educate all those chauffeurs?}{Gottlieb Daimler}
A decade ago, we cited the above quote to characterize the state of software development~\cite{MarSte2010}.
History has shown that it is not necessary to make everyone a chauffeur in order to drive a car.
\acf{lcd}\acused{lcd} and No-Code Development, also called programming-less development~\cite{BFKLNN2016}, follow exactly this line of thinking: People can be developers without being a programmer.
\ac{lcd} expresses the search for sufficient \textbf{simplicity} to allow people with little or no programming knowledge, such as application experts, to contribute {\em directly} to the development process.
To achieve this, \acfp{dsl}\acused{dsl} are created.
These languages aim for simplicity through abstraction, but must also be \textbf{accurate} enough to be understood by a machine.
Considering Thorngate's postulate of commensurate complexity that ``at any one time one can achieve only two of the three meta-theoretical virtues of Generality, Accuracy
and Simplicity.''~\cite{Weick1999}, this means that a low-code approach using \acp{dsl} must sacrifice some \textbf{generality} in order to be accurate and yet simple.

\label{subsec:cinco-language-workbench}

Every craftsman relies on the power of specialized tools.
Carpenters use special knives, painters use special brushes, and tailors use scissors, needles, and thread.
Production lines, common in automotive manufacturing take this paradigm to another level, where specialized tasks for specific purposes are seamlessly coordinated in a machine-assisted workflow.
\ac{lde} aims to apply this idea to systematic software development.
Similar to car manufacturing, where different tasks are coordinated to refine the \emph{one thing}, the car, we provide purpose-specific \acp{ime} for the stakeholders involved.
It is important for such a product line approach to be economical. 
The cost of building the product line must be small compared to the intended gain.

Our \cinco language workbench enables the design and automatic generation of economical \acp{dsl}.
It allows users to create distinctive graphical modeling tools for different platforms.
All of the applications presented in this work are based on \cinco. 
The presentation of these projects is intended to illustrate the power of domain specificity, to bridge the gap between construction engineers, data analysts and computer scientists, and to give students the means to easily experience the power of formal modeling.
\dime{}, the largest \cinco product described in \autoref{sec:dime}, is intended to illustrate the breadth of the \ac{lde}.
It is used in cross-department projects that include industrial cooperation involving historians and robotics experts.
\dime{}'s flagship application, Equinocs, is Springer Nature's new editorial service.
It supports program committees from the receipt of submissions through the peer review process to the final delivery of the built proceedings to Springer Nature.
Equinocs itself, as well as the subsequent deployments at Springer Nature, are generated fully automatically from graphical models using an automatically generated \ac{ci/cd} pipeline.
The only manual code written is for the ever growing service library.

\paragraph{Contributions}
\label{Purpose-and-Contribution}

We illustrate how domain-specific graphical modeling enables application experts with different levels of expertise to cooperatively design and then deploy complex applications using their individual, purpose-specific modeling language.
Achieving such a shift of modeling to the application expert requires adequate technology to transform notations popular in a domain into full-fledged modeling languages that support complete code generation. 
Previous language workbenches provide similar technologies, but typically at a high cost of manual coding.
For our service-oriented approach, we rely on \cinco, our meta-tooling suite with a strong focus on simplicity of language development~\cite{NaLyKS2017}.  
In addition, the seamless integration of the multiple languages involved in a complex development process is a matter of \ac{lde}~\cite{StGoNM2019}, where the \ac{ota}~\cite{MarSte2009} provides a
modeling structure that ensures consistent alignment of artifacts modeled across languages.

\begin{figure}[tb]
	\centering
	\includegraphics[width=0.95\textwidth]{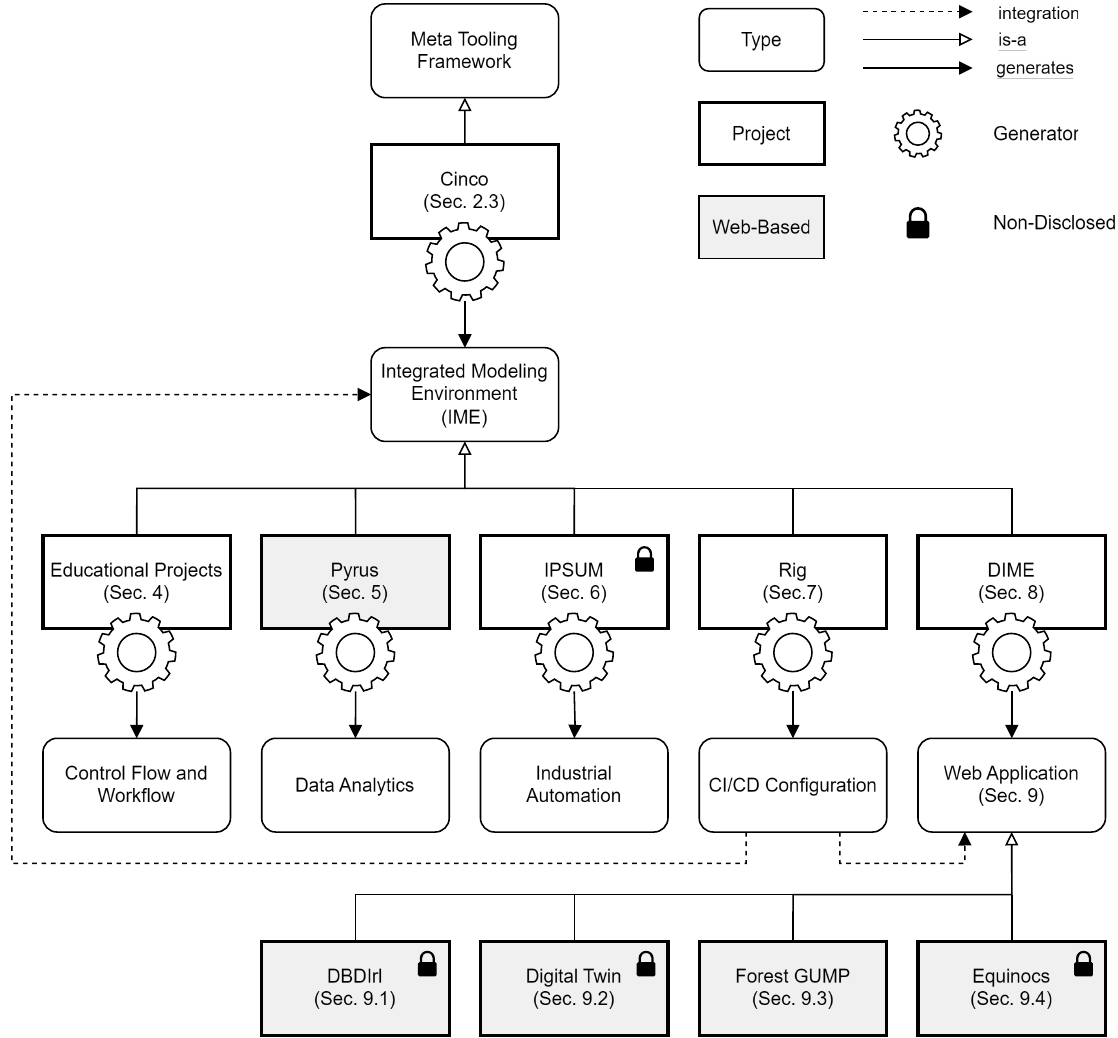}
	\caption{The \acs{lde} landscape discussed in this paper}
	\label{fig:overview-ontology}
\end{figure}

We illustrate the impact of \ac{lde} as a low-code paradigm with concrete implementations.
All of them are fully automatically generated from models. 
\autoref{fig:overview-ontology} summarizes the structure and relationships of the considered tool landscape.
In the figure, the fourth row shows typical \cinco products. 
These \acp{dsl} range from simple educational languages, explained in~\autoref{sec:others}), via data analytics, industrial automation and \ac{ci/cd} configuration, cf. Sec.~\ref{sec:ipsum} -- Sec.~\ref{sec:rig}, to the complex \dime application for developing browser-based applications in~\autoref{sec:dime}.
The sixth row in the figure shows four products created with \dime. 
They are all browser-based applications, ranging from smaller academic products to Equinocs.
The required \ac{ci/cd} pipeline was automatically generated from models in the Rig.
As indicated in the figure, almost all projects are open source.
Interested readers can consult our website\footnote{\url{https://scce.info}} and repositories on GitLab\footnote{\url{https://gitlab.com/scce}} for more information and project insights.

\paragraph{Outline}

The discussion in this paper is organized as shown in~\autoref{fig:overview-ontology}.
\autoref{sec:background} discusses the foundations of \ac{lde} and gives further background information.
\autoref{sec:related-work} describes related work.
The conceptual background is provided by the three paradigms \ac{lde}, \ac{ota}, and \ac{lop} mentioned above, which are explained in more detail in the following section. 
We consider these three paradigms as the key to bringing the technology closer to the application experts by providing them with an interface that resembles their expertise and mindset.
\autoref{sec:others} -- \autoref{sec:dime} then describe in detail concrete projects from different domains that were created with \cinco.


\section{LDE Foundations}
\label{sec:background}

This section concerns mindsets, views, and experiences where the choice of notation made a difference: languages impose mindsets.
\autoref{sec:history} recapitulates the events that lead up to the \ac{lde} that we have today.
\autoref{sec:language-driven-engineering} then describes the \acs{lde} fundamentals and the \autoref{sec:the-one-thing-approach} explains the \ac{ota} key enabler.

\subsection{History}
\label{sec:history}
\label{subsec:roots}

We began to observe the tremendous impact of changing perspective from a computational \textbf{how}  to a
logic-based \textbf{what}.
Often, supposedly critical problems fell apart after being reformulated in an appropriate way.

The next observation was that one has to speak the language of the people in a particular domain if one wants to convince them.
Typically, we computer scientists think we can take the lead because we are the only ones who can ``talk to computers''.
The low-code approach tries to enable the application experts to participate in software development by drastically simplifying the coding.
We consider such simplification alone as insufficient, because it does not necessarily reflect the mindset behind the application under consideration.
If it is not done in a mindset-aware fashion, even low coding is still coding.
We first experienced the power of this approach with the \acfp{slg} we used in a project with Siemens/Nixdorf.
In that project, part of the success was due to the formation of a successful mindset. 
Our industry partners began to use well-designed templates for temporal formulas to formulate critical requirements for their telecommunications infrastructure.
At a second time, this change to the \emph{what perspective} became particularly clear in the IPSUM project, which will be discussed in more detail in \autoref{sec:ipsum}.
The mindsets of the stakeholders were very different, but each of them expressed their views in their customary notation, e.g. in the form of a \ac{pid}, served to provide the alignment required for the \ac{ota}.
Here, five domain-specific languages directly reflected the mindset of the industry partner.

Our first direct experience demonstrating the power of \ac{dsl}-based mindsets came in 1991 with an attempt to prove the optimality of a partial redundancy elimination algorithm proposed by Morel and Renvoise~\cite{MorRen1979}.
Instead of thinking in terms of fixpoint computations, as was common at the time, thinking in terms of temporal logic properties was a radical change in mindset.
This led to shorter proofs and later allowed us to solve the two related problems of optimal reduction of register pressure~\cite{KnRuSt1992,KnRuSt1993} and elimination of all partial redundancies~\cite{Steffe1996}.
The former solution is now the standard for optimizing compilers.

\begin{figure}[tb]
	\centering
	\includegraphics[width=\textwidth]{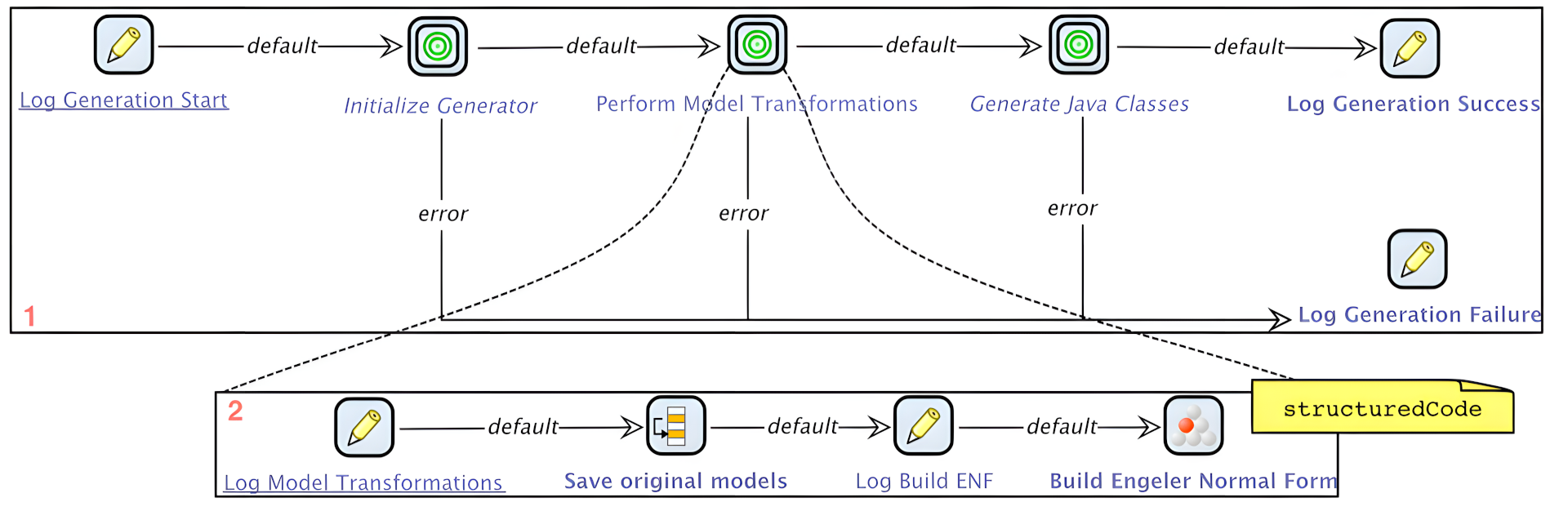}
	\caption{Java Class Generator of the Genesys framework (reprint of~\cite{Jorges2013})}
	\label{fig:javaClassGeneratorTopM2M}
\end{figure}

This mindset also gave rise to the idea of introducing code generators~\cite{Steffe1991,Steffe1993a} that could benefit from the considered domain.
Full code generation became a central goal in all further developments.
It was supported by a dedicated, graphical \ac{ime} called Genesys~\cite{Joerges2011}.
\autoref{fig:javaClassGeneratorTopM2M} shows a Genesys-based code generator model.
The idea of using reusable standalone components, later called \acp{sib} in the ITU-T Standard~\cite{ITU1993}, which can be easily recombined due to the simplicity of their interfaces, was motivated by the rapidly growing library of special commands for the
Concurrency Workbench~\cite{ClPaSt1993}.
\acp{sib} provided an adequate 
level of abstraction to introduce model checking and model synthesis~\cite{StMaFr1993,MaMKIS2009}.

This was the basis of  the \ac{eti}~\cite{StMaBr1997}, which was later specialized to Bio-jETI~\cite{LaNaMS2010a}, a \ac{sib}-based graphical \ac{ime} for modeling scientific workflows in bioinformatics.
Bio-jETI was a service-oriented, low-code environment. 
Users could graphically model their workflows (processes), which could then be enacted without any programming.

\begin{figure}[t]
	\centering
	\includegraphics[width=\textwidth]{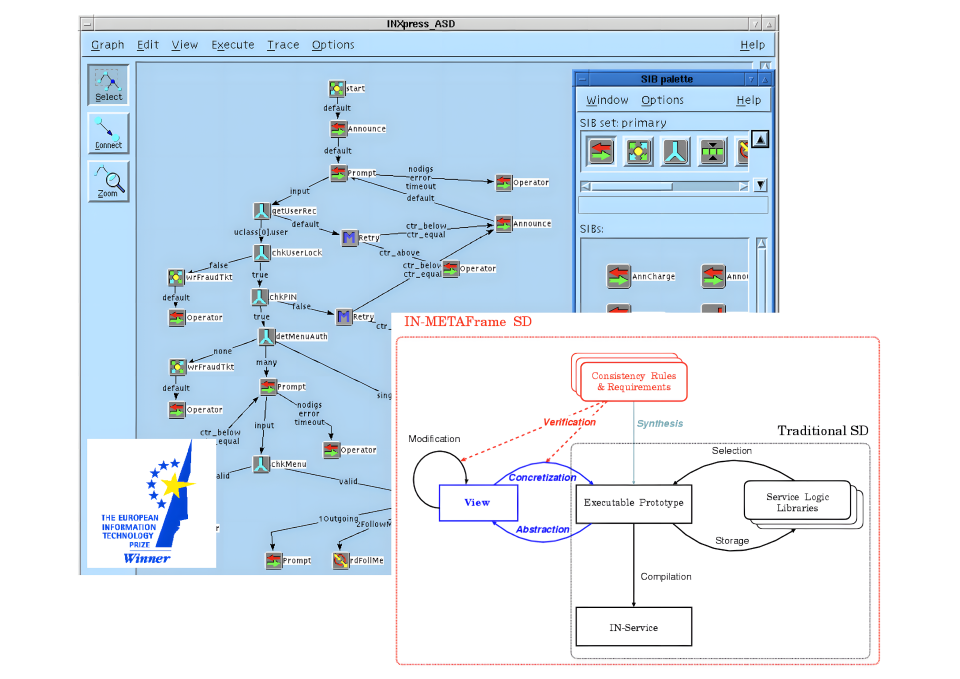}
	\caption{The 1996 Siemens Nixdorf INXpress SDE and an overview of its conceptual backbone.}
	\label{fig:metaframe-collage}
\end{figure}

However, the greatest influence on our service-oriented development approach came from a successful industrial cooperation that resulted in the Siemens Nixdorf INXpress Service Definition Environment for Intelligent Networks ~\cite{SMCBNR1996,StMCBR1996} in the mid-1990s.
The \ac{sde}, based on the \ac{sib} concept combined with taxonomic classification and model checking, was evaluated to reduce the time to market by a factor of five and received an ITEA~award in~1996.
\autoref{fig:metaframe-collage} shows a screenshot of the \ac{sde} and a conceptual diagram illustrating the underlying constraint-based approach. 
In retrospect, the \ac{sde} can be considered one of the first low-code environments.
Telecommunication experts could graphically combine \acp{sib} into \acp{slg}, which were continuously controlled by model checking.
In fact, even we, the \ac{sde} developers, did not have access to the implementation code of the actual \acp{sib}, which was considered proprietary. 
Our domain-specific modeling environment \dime{} for web-based applications, which will be discussed in detail in \autoref{sec:dime}, can be seen as the sixth generation of our original \ac{sde}. 

The simple service-oriented definition and exchange of functionality turned out to be a good way to communicate between the stakeholders~\cite{MaStRe2005,MaStRe2006}.
All stakeholders were working on the same artifact, which we later called the \emph{one thing}~\cite{SteNar2007,MarSte2009} but at the appropriate level of abstraction that supported their mindset and was defined by the underlying service hierarchy.
The underlying philosophy of full code generation was the basis for \acf{jabc}, a fully fledged low-code environment~\cite{KuJoMa2009,MarSte2009a}.
In addition, its model checking and model synthesis facilities provided dedicated support for controlling evolution at the logical, declarative level and for establishing product lines via their behavioral (temporal) properties~\cite{JoeLMSS2012,LaNaSc2013}.
The latest version of \ac{jabc} supports even higher-order services~\cite{NeuSte2013b,NeStMa2013}.

The final enabling step for \ac{lde} was to move from \acp{dsl} defined by taxonomically organized service libraries to \ac{cinco}, our framework for metamodel-based generation of graphical editors and full-featured \acp{ime}~\cite{NaLyKS2017,Naujokat2017}.  
This step had two major impacts:
\begin{itemize}
	\item First, the move from \ac{jabc} to \ac{cinco}'s web-\ac{ime} \dime was a move from a process model with manual handling of data, GUI and deployment in \ac{jabc} to an integrated modeling environment where dedicated environments for data, process, GUI, and CI/CD pipelines put these dimensions on equal grounds.
	These four \acp{dsl} provide purpose-specific views that allow the corresponding experts to model their part in their own mindsets. 
	The first three \acp{dsl} are deeply integrated into \dime, meaning that the corresponding environment are part of the \dime{} \ac{ime}. 
	The integration of the \ac{dsl} for CI/CD is shallow, meaning that
	\dime{} simply integrates the generated CI/CD pipeline, which will be illustrated further in \autoref{sec:rig}.
	This move was only possible by basing our entire \ac{ime} generation on \cinco{} because of the very different nature of the involved \acp{dsl}.
	
	\item 
	Second, the other projects listed in row four of \autoref{fig:overview-ontology} required the flexibility of \cinco{}-based \ac{ime} generation. 
	For example, this was needed to implement the token-driven PetriNet semantics, to enable data flow-driven execution semantics in Pyrus, and to enhance customer languages like \ac{pid} and hardware diagrams in IPSUM.
\end{itemize}

\subsection{Language-Driven Engineering}
\label{subsec:language-driven-engineering}\label{sec:language-driven-engineering}

\acused{psl}
\ac{lde} addresses the need to provide low-code environments for the development of industrial-scale solutions, so that multiple stakeholders with different backgrounds and expertise can directly participate in the development process using Purpose-Specific Languages (PSLs) tailored to their specific mindsets.
This means that interdisciplinary teamwork in an application domain typically requires multiple \acp{psl} to support the different needs of each team member.
Thus, in contrast to \acp{dsl}~\cite{StGoNM2019}, \acp{psl} are stakeholder-specific and tailored to achieve a particular goal; for example, in a development team, GUI designers, process engineers, quality assurance, and operations teams are supported by different \acp{psl}.

\acp{psl} are essential for efficient inter-stakeholder collaboration and communication, as they provide each team member with appropriate language support, as pointed out in \autoref{subsec:roots}.
The key point of LDE is to consider language provision itself as a service.
Adequately integrated PSLs can provide support at different levels, e.g.:
\begin{itemize}
	\item \emph{\acs{ime} support}, so that stakeholders can model the user-level functionality, which is then included in a service-oriented fashion, in a \ac{psl}.
	Examples of such \acp{psl} are dedicated (graphical) query languages, languages for data analysis, e.g. as supported by Pyrus, as illustrated in \autoref{sec:pyrus}, but also languages derived from graphical languages traditionally used in application domains, such as network layouts, workflow graphs, or \ac{pid}, as shown in \autoref{sec:ipsum}.

	\item \emph{\acs{ime} enhancement} via languages for modeling functionality that is then provided by the tools, e.g., via plug-ins. 
	Such plug-ins provide quality assurance in the form of model checking and testing, various code generators, and platform-specific \ac{ci/cd} pipelines.
	
	\item \emph{Tailored interaction facilities} for specific users, such as simple configuration languages (cf.\ the \ac{ci/cd} pipelines presented in~\autoref{sec:rig}), canvases for business modeling, and spreadsheets or data flow graphs for data analysis.
\end{itemize}

We also distinguish between deep and shallow language integration. 
This is a technically important distinction that tends to become fuzzy in the context of metamodel-driven language workbenches like \ac{cinco}. 

\begin{itemize}
	\item In \emph{deep language integration}, the special-purpose \acs{ime} are integrated as a service into the global development \ac{ime}. 
	The data, process, and \ac{gui} language integration in \dime{} (see \autoref{sec:dime}) are typical examples of deep integration.
	
	\item \emph{Shallow language integration} provides separate special-purpose \acs{ime} that produce artifacts for the component library of the development IDE. 
	In this case, the artifacts generated by the special purpose \acs{ime} are integrated as a service.
	From the meta-level perspective, a \ac{psl} with an \ac{ime} that is generated from any meta-model within the \ac{cinco} meta-tooling suite can be considered both as deeply integrated and as a special-purpose IDE resulting from the shallow integration of \ac{cinco}.
	For example, for an artifact that is itself an \ac{ime}, in Pyrus we decided to do a shallow integration of the data \ac{psl} and simply provide a standard data model as a service.
	This decision lowered the entry barrier and led to higher acceptance by the users. 
	In (rare) cases where the standard data model is not sufficient, one can use \ac{lde} to easily model and then (shallow) integrate an extended data model.

\end{itemize}
Whatever the perspective, the practical difference between shallow and deep integration provides a very powerful \ac{lde} discipline for tailoring \acp{ime}~\cite{StGoNM2019}.
Deep integration increases the conceptual complexity of the original \ac{psl}, while shallow integration provides additional external services to the users of the development \ac{psl}. 
For example, data analysis processes modeled with Pyrus.
\subsection{The One Thing Approach}
\label{subsec:the-one-thing-approach}\label{sec:the-one-thing-approach}

The key to successful \ac{lde} is the \ac{ota}.
It combines the simplicity of waterfall development with maximum agility~\cite{MarSte2009} to ensure consistency.
\ac{ota} is characterized by viewing the entire development process as a cooperative hierarchical, and interactive decision-making process organized by building and refining one comprehensive model, the \emph{One Thing}. 
This approach allows each stakeholder, including the sponsor, the principal, and application experts, to make decisions in terms of constraints on an ongoing basis using their dedicated \acp{psl}.
Conceptually, each evolutionary step is viewed as a transformation or reification of this set of constraints,  which typically comprises many modeling aspects and embodies multiple forms of knowledge.
Dedicated \acp{psl} and their associated views highlight open requirements and potential conflicts, and support goal-oriented and stakeholder-specific design decisions.
A clear chain of command, authority, and responsibility assigns levels of priority to the stakeholder decisions according to their roles.
This chain is organized to support intent-driven development.
Typically, this means that the sponsor's or the principal's constraints have the highest priority and that the priority decreases as the level of technical detail increases.

\ac{lde} provides means to check for conflicts during the decision making process.
The One Thing is the single source of truth and every other artifact is automatically generated from that source. 
This is a central point that prevents any form of round trip engineering. 
In fact, change requests can only be entered into the One Thing and by a person with the appropriate role. 
Any consequences for the lower levels are automatically enforced, i.e. communicated to corresponding stakeholders, via dedicated views~\cite{JoeLMSS2012}.
The success of this approach is based on the interplay of the following three concepts:
\begin{itemize}
	\item
	\emph{Language-oriented decomposition} that breaks down complex development tasks into specific subtasks, ideally handled by dedicated (domain) experts using a \ac{psl}.
	This decomposition works the better the closer the modeling language resembles (drawing) patterns that already have a tradition in the specific domain.
	
	\item
	\emph{Global alignment through a global model}, with simultaneous separation of concerns through specialized \acp{psl} for each subtask. This holistic approach of bounded contexts can be seen as a solution for \ac{ddd}~\cite{Evans2003}	where bounded contexts are a central aspect to minimize complexity while increasing flexibility, especially when each stakeholder can use \acp{psl} specially designed for their needs.
	Considering the different \acp{psl} for data, process and GUI modeling of \dime again, the incrementally growing data model that is built during system development is the backbone of the One Thing. 
	Processes are defined in relation to this data model and also refer to each other.
	Each of the modeling editors involved provides the necessary views of the artifacts created by the other modeling editors.
	For example, the process model and the GUI model in \dime{} contain customized views of the data model (see more details in \autoref{fig:dime-process-data-gui}).
	
	\item
	\emph{Constraint-based guidance, validation, and control} manage complexity through automated assistance because the One Thing is multifaceted and hierarchical.
	For this reason, application constraints are formulated and continuously validated at the meta level, the development level, and at runtime.
	For example, our \ac{ci/cd} pipeline, which seamlessly connects the meta-modeling level with the production	level, automatically checks all constraints, ensuring that violating systems do not reach the production platform.
\end{itemize}
%


\section{Related Work}
\label{sec:related-work}
\label{subsec:relwork}

Two properties are characteristic for \ac{lde} and its corresponding \acp{ime}.
First, \ac{lde} aims to enable all stakeholders to co-develop software without programming.
Second, it explicitly supports multi-\ac{psl}--based collaboration of the individual stakeholders.
The related work can be divided into approaches that address these two characteristics.

\paragraph{Languages for Non-Programmers}

Fowler, who coined the term \emph{language workbenches}~\cite{wwwFowler2005}, characterizes the role of (textual) \acp{dsl} such that it is not that domain experts write the \acp{dsl} themselves, but that they can read them and thus understand what the system thinks it is doing~\cite{FowPar2011}.
Low-code modeling approaches that rely on language workbenches, such as MetaEdit+~\cite{wwwmetacase15,KelTol2008}, the Eclipse Modeling Project~\cite{Gronback2009}, Sirius~\cite{wwwsirius14}, and (Web) GME~\cite{LMBKGT2001}, explicitly address this mindset issue. 
They allow the language to be tailored to the skills and needs of the target domain experts. 
In addition, several graphical languages have become successful in dedicated application domains, such as MatLab/Simulink\footnote{\url{https://www.mathworks.com/products/simulink}}, ladder diagrams~\cite{JohTie2010}, and Modelica~\cite{Fritzo2004}.
Recently, large software companies have been pushing their low-code platforms, such as Google's AppSheet\footnote{\url{https://www.appsheet.com}} or Microsoft Power Apps\footnote{\url{https://powerapps.microsoft.com}}.
They provide a manually built development solution that resembles a spreadsheet-oriented mindset. 
The tradeoff between simplicity and generality is implicit: Simplicity is given whenever the spreadsheet-oriented mindset is appropriate.
Other prominent examples, such as Mendix\footnote{\url{https://www.mendix.com}}, Bubble\footnote{\url{https://bubble.io}}, or Salesforce Lightning\footnote{\url{https://www.salesforce.com/campaign/lightning}}, focus on (graphically) specifying workflows in a process model or flow graph style using manually programmed, application-specific activity blocks. 
These approaches address situations where a process-oriented combination of predefined building blocks is sufficient.
This was the mindset behind the \ac{jabc}~\cite{StMaNa2006} before we started developing \cinco{} in an industrial project, which is the basis of the low-code approach presented in this article. 
Similar to the spreadsheet-oriented mindset, the adequacy of solutions depends heavily on a good fit between the mindset of the toolset and the application being addressed.
The understanding that application experts should be addressed with graphical notations is also shared by the developers of the \textsc{Kieler} framework~\cite{FuhHan2010}.
They provide means to automatically generate domain-specific graphical views for textual \acp{dsl} realized in the Eclipse modeling context.
While the \textsc{kieler} framework is indeed mature and powerful---so much that its layout engine is now generalized as the Eclipse project \emph{Eclipse Layout Kernel} (ELK)\footnote{\url{https://eclipse.dev/elk}}---its primary goal is to provide views to better communicate with non-programmers, while creating the actual (often textual) models still requires programmers or highly technically experienced domain experts.
We therefore focus on graphical \acp{psl}.

Prominent frameworks for graphical modeling language development include MetaEdit+~\cite{KeLyRo1996}, GME~\cite{LMBKGT2001}, Marama~\cite{GHLAHL2013}, Pounamu~\cite{ZhGrHo2004}, and DeVIL~\cite{ScCrKa2008}.
These powerful frameworks are designed to generate graphical \acp{ide} for specified graphical \acp{dsl} including corresponding code generators.
However, they do not address the aspect of coordinating stakeholders with different mindsets in a cooperative fashion.
The same applies to the Eclipse\footnote{\url{https://eclipse.org}} modeling ecosystem~\cite{Gronback2009} with its \acp{rcp}~\cite{McLeAn2010} and \ac{emf}~\cite{StBuPM2008}.
Although there is extended support for textual \acp{dsl} in Eclipse, e.g. using the Xtext~\footnote{\url{https://eclipse.dev/Xtext}} framework, the creation of graphical \acp{dsl}, usually using frameworks such as GMF~\footnote{\url{https://eclipse.org/gmf-tooling}} or the Epsilon project~\cite{KRAPPB2010}, becomes tedious.

Lowcomote~\cite{tisi2019lowcomote} is a recently launched research project that explores current challenges in low-code development platforms.
One of the goals of the project is to scale low-code approaches beyond simple \ac{crud} applications into domain-specific areas of interest, such as data science of \ac{iot}.
The project also aims to address the increasing fragmentation among low-code platforms with open standards and programming models that enable greater interoperability.
Furthermore, the integration of heterogeneous models from different disciplines is an explicit goal of the project.
Compared to our research, the Lowcomote project is relatively young, having only started in September 2019.
We have been working on the vision of \ac{lde} for more than two decades, developing a comprehensive metamodeling framework for graphical \acp{psl}, including model validation, full code generation, and language implementations for many domains.
Nevertheless, there are interesting published results that are relevant to our research in the sense that domain-specific \ac{mde} has been applied to the \ac{iot} domain~\cite{DBLP:journals/corr/abs-2105-14136}.
DevOpsML~\cite{10.1145/3417990.3420203}, also developed within the Lowcomote project, uses \ac{mde} for the purpose of modeling DevOps workflows.
It attempts to capture the bigger picture of the complete DevOps workflow, including development, build, test, deployment, and operations.
Software processes and DevOps platforms are modeled graphically, and then woven together using a link model that connects the requirements of the software process with the capabilities of the DevOps platforms.
DevOpsML focuses on a higher level, without a clear distinction between DevOps and \ac{ci/cd}.
Rig considers \ac{ci/cd} as the central building block of DevOps and provides a graphical \ac{psl} as a comprehensive and self-contained approach that can be integrated in a shallow way into any DevOps process.
It has been successfully integrated in industrial projects, such as Springer's Equinocs, and generates complete \ac{ci/cd} pipelines.

The modeling language CHESSIoT for \ac{iot} systems was developed on top of the CHESS tool~\cite{modelsward21} to improve the safety of \ac{iot} systems and to facilitate validation through model checking.
It provides metamodels for system, software, hardware, and operational aspects.
As model checking and validation are widely applied in our \ac{lde} research, our goals are similar in this regard.
While our research is focused on \acp{psl}, CHESSIoT is dedicated to the entire industrial \ac{iot} domain without further specialization for a specific purpose. 
In addition, full code generation is an essential part of the \ac{lde} process, whereas CHESSIoT is targeting the ThingML~\cite{10.1145/2976767.2976812} modeling language, which has not yet been implemented.

In the area of model-driven web application development, Bozzon \etal~\cite{DBLP:conf/icwe/BozzonCFT08} explore how model-driven approaches can support the transition to Web 2.0.
They use models to capture important social and technological aspects of this transition.
The beContent project~\cite{10.1007/978-3-642-02818-2_52} defines a programming system where graphical models and textual programming of a web application are synchronized in a round-tripping process.
In contrast, \dime{} eliminates the round-tripping process by implementing full code generation.

\paragraph{Language-Driven Development}

The Racket team's~\cite{FFFKBM2018} \ac{lop} approach is aligned with \ac{lde} in terms of explicit support for multi-\ac{dsl}.
In fact, \ac{lop} advocates multiple cooperating languages for a project, has a feature called \ac{ffi} (Foreign-Function Interface) similar to our notion of native services, and uses a meta-language \emph{syntax parser} to define languages.
However, there are clear conceptual differences that limit cooperation with non-programmers. 
\ac{lop} relies on internal \acp{dsl}, called \acp{edsl}, based on the single base language Racket\footnote{\url{https://racket-lang.org}}, which is a successor to Lisp~\cite{Weissm1967}.
Discouragingly, even exemplary \ac{edsl} code must be annotated with simple graphical notations for readability to communicate with non-programmers.~\cite{AnChFe2017}

Another prominent approach to model-driven development is projectional editing~\cite{VoSiBK2014}, such as provided by JetBrains' \ac{mps}\footnote{\url{https://www.jetbrains.com/mps}}.
This approach is mainly aimed at programmers who are proficient in different (programming) languages.
\acp{psl} relate to projectional editors, the \emph{one thing} to \ac{ast}, and our constraints to the wealth of constraints that can be imposed on \acp{ast}. 
The difference is that \ac{lde} is technologically free-style due to \cinco~s service-oriented fashion, as long as consistency is maintained. 
This means that an adequate extension of a technology usually also imposes the need for a \acs{psl} that defines the corresponding consistency constraints. 
For example, adding model checking to the \dime{} landscape required a \ac{psl} for transforming \dime{} models into verifiable models that can be realized using the transformation approach.~\cite{KoLyNS2021}

The ROCCO tool~\cite{EuGENia2020} has similar goals. 
It makes graphical modeling languages based on Eclipse EuGENia\footnote{\url{https://www.eclipse.org/epsilon/doc/eugenia}} accessible in a web-based low-code development platform. 
ROCCO performs the migration by generating diagram models for evaluation in the Psi Engine~\cite{CHAVARRIAGA2017133}, a web-based low-code environment. 
Compared to Pyro, ROCCO does not provide collaborative editing of the diagram models.


\section{Educational Projects}
\label{sec:others}
\cinco provides students with a tangible experience with formal modeling.
The following subsections present two prominent examples, first, a \ac{psl} for classical control structures in \autoref{sec:classical-control-structures}, and second an intuitive tool for simple point-and-click games in \autoref{sec:webstory}.

\subsection{Classical Control Structures}
\label{sec:classical-control-structures}

\begin{figure}[tb]
	\includegraphics[width=0.66\textwidth]{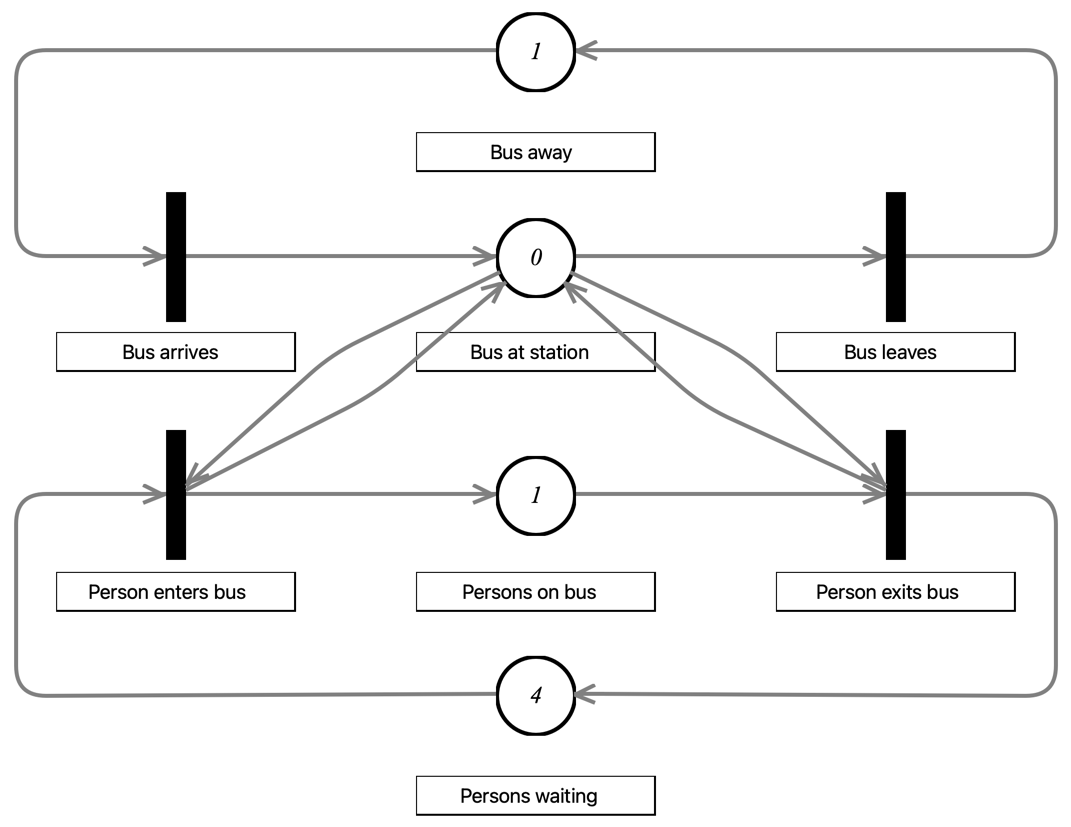}
	\centering
	\caption{Example of a bus station modeled in a \cinco \textsc{Petri Net} tool}
	\label{fig:petrinet}
\end{figure}

\begin{figure}[tb]
	\begin{subfigure}[c]{0.80\linewidth}
	\includegraphics[width=\textwidth]{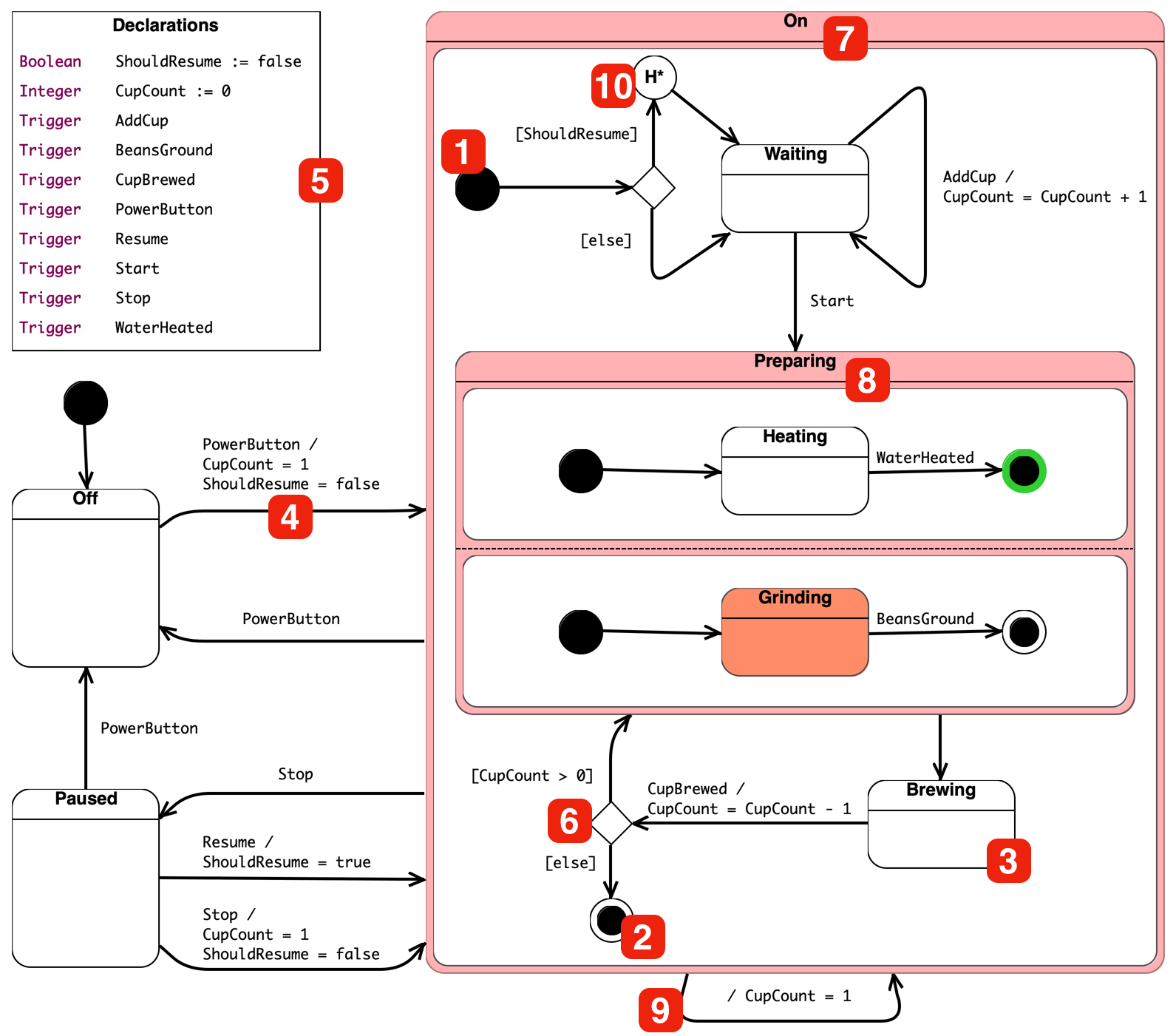}
	\centering
	\caption{Example of a coffee machine modeled in a \cinco \textsc{statechart} tool}
	\label{fig:statechart-coffee-marked}
	\end{subfigure}

	\begin{subfigure}[c]{0.5\linewidth}
	\includegraphics[width=\textwidth]{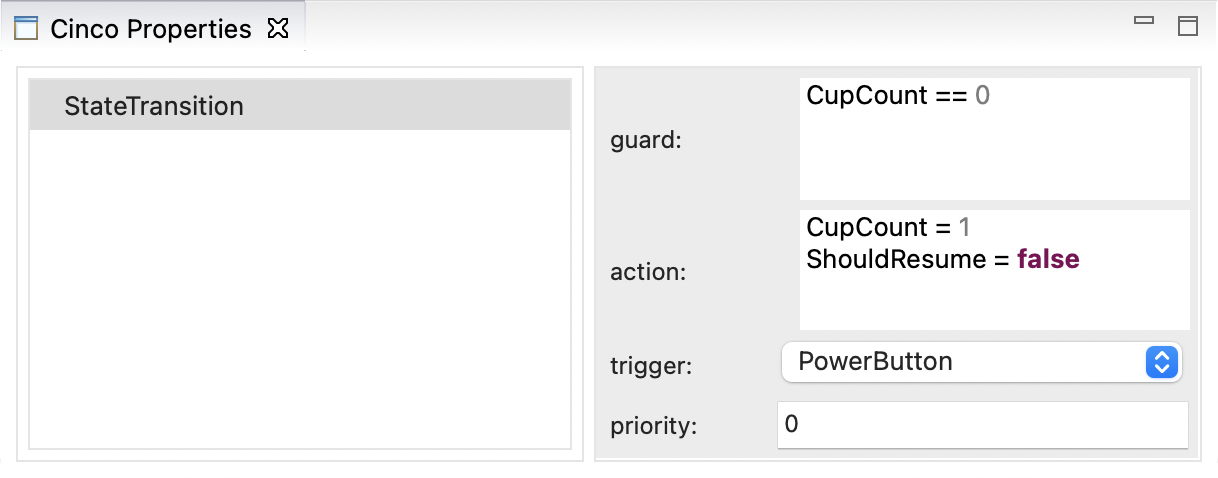}
	\centering
	\caption{The property view of a  \textsc{State Chart} transition.}
	\label{fig:statechart-transition-properties}
\end{subfigure}
	\caption{Example of a coffee machine modeled in a \cinco \textsc{statechart} tool}
	\label{fig:statecharts}
\end{figure}

A variety of diagram types are commonly used in software development and computer science. 
There are tools for modeling complex systems and visualizing processes, behaviors, and interactions. 
Their graphical nature makes them a natural use case for graphical \acp{psl} and therefore reasonable examples for \cinco products. 
The following two figures show two examples of models modeled with a tool generated with \cinco.
First, a \textsc{Petri Net} in \autoref{fig:petrinet}, and second a \textsc{statechart} in \autoref{fig:statecharts}.
These projects automatically provide the corresponding editors and checkers and may be used to teach how to use these specific diagram types. 
They provide a responsive hands-on experience, including model validation and the ability to run simulations. 
Furthermore, they also demonstrate core features of \cinco and how to implement them.
Hence, \cinco product developers can use them as reference implementations for creating their own \acp{psl}.

Focusing on the \textsc{statechart} in \autoref{fig:statechart-coffee-marked}, which allows the creation and simulation of finite-state machines, the corresponding \ac{psl} comprises a subset of the \ac{uml} state machine specification.
Basic model elements, such as \emph{start} (1), \emph{end} (2), and \emph{state} (3) nodes, as well as \emph{transitions}, (4) are available. 
Variables, which can be boolean or integer, and triggers are defined in the declarations container (5). 
During simulation, their current values are displayed in a table view. 
Triggers symbolize external events or inputs to the modeled system. 
They can be activated using buttons in the simulation interface. 
The properties of transitions, shown as an example in \autoref{fig:statechart-transition-properties}, define if and when a transition can be activated. 
They can refer to both variables and triggers. 
If a transition's source node is active (highlighted in orange), its guard expression evaluates to \emph{true}, and the trigger is activated, then the transition becomes active and its action is executed. 
Afterwards, the target node of the transition becomes active. 
The guard and action text fields are each editor views, that use textual \acp{dsl} powered by \textsc{Xtext}~\cite{EEKZMH2012} for boolean expressions and variable assignments. 
They support syntax highlighting, syntax validation, and code completion. 
In addition to these basic elements, there are also a stateless decision nodes (6) that use guard expressions to branch the control flows.

Statecharts also include rather powerful concepts, such as hierarchical (7) and concurrent (8) states. 
Hierarchical states allow the user to group several states together. 
When the \emph{end node} inside a nested state is activated, the default transition---that is, the only transition without an assigned trigger (9)---is activated. 
The execution of a nested state can be interrupted by using an outgoing transition with an assigned trigger. 
For example, triggering a {\em Stop} in \autoref{fig:statechart-coffee-marked} will always transition from the {\em On} state to the {\em Paused} state. 
All internal states are deactivated. 
However, the last active states are saved by a \emph{history node} (10) (if present). 
Activating such a node will restore the last active states. 

The aforementioned concurrent states (8) are a special case of the hierarchical states. 
Their two or more regions are states that are executed concurrently. 
Like the nested states, they can be interrupted and use history nodes to restore their active states. 
The difference is that each region's end node must be active in order to activate the outgoing default transition. 
The {\em Preparing} state in \autoref{fig:statechart-coffee-marked} has two regions. 
The first one is finished, as its end node is already active (green). 
The second region is not yet finished, because the coffee beans are still being ground.

\subsection{The \webstory Language}
\label{sec:webstory}

The \webstory language~\cite{10.1007/978-3-030-03418-4_31} is a simple example for \cinco.
It is a graphical \ac{psl} for easy authoring of basic Point\&Click adventure games, and comes with a modeling tool to assist users creating such games.
Modeled \emph{WebStories} are generated into source code based on web technologies, so the games are fully functional websites that can be executed by a browser.
Following the {\em storyboard} concept, a \webstory basically describes the flow from one screen to the next.
Each screen primarily displays a background image and contains one or more clickable areas.
When playing the game, the player moves from one screen to another by clicking on certain areas of the image to reach a predefined goal.
Between screens, a trigger can change the state of the game, e.g., picking up a key.
Target screens may be unreachable until the game is in a certain state, e.g., when the correct key has been picked up.

\begin{figure}[ht]
	\centering
	\includegraphics[width=\textwidth]{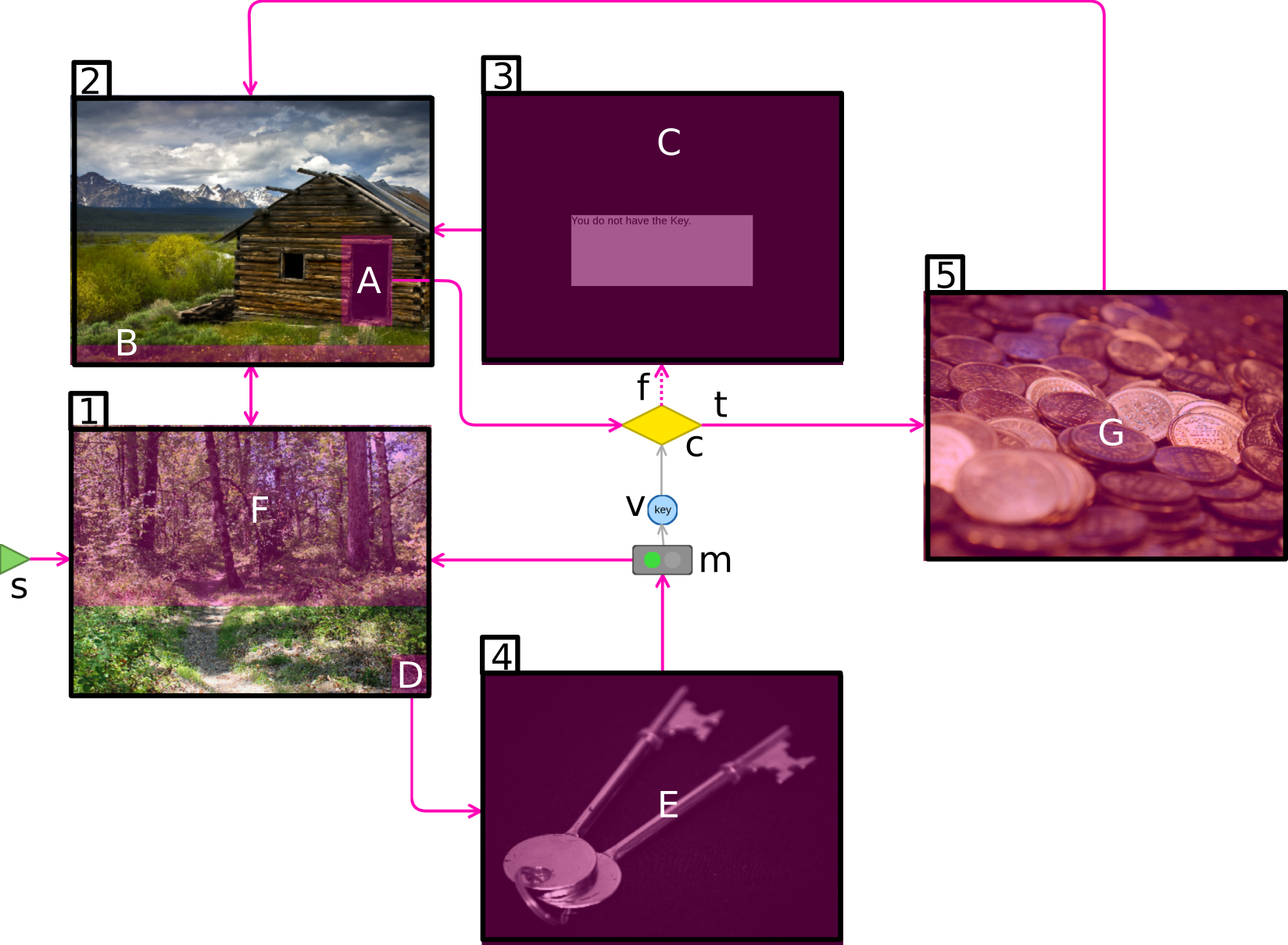}
	\caption{An exemplary \webstory model, reprint of~\cite{10.1007/978-3-030-03418-4_31}\protect\footnotemark}
	\label{fig:webstorysimple}
\end{figure}
\footnotetext{Images by Iain Watson, David Geitgey Sierralupe, Charles Knowles, and Adam Dachis from \url{https://www.flickr.com}}

\autoref{fig:webstorysimple} shows an example \emph{WebStory} where the goal of the players is to find a treasure.
The treasure can only be obtained if the player finds the required key.
The \webstory uses several language elements alongside the basic \emph{screens}~\textsf{1-5}.
The \emph{click areas} \textsf{A-G} connect two screens by purple directed control flow edges.
The \emph{start marker} \textsf{s} defines the entry point for the player.
Triggers serve as \emph{variable modifiers} \textsf{m}. 
The boolean \emph{variable} \textsf{v} marks the possession of the key, which is then tested by the \emph{condition} \textsf{c}.
Conditions and variable modifiers are connected to variables by grey data edges.
The values of all variables define the state of the game.
In this example, the \textsf{key} variable holds the information whether the player has the key or not.
The value of the key variable is potentially changed by the variable modifiers and evaluated by the condition.
A \emph{true} edge \textsf{t} links to the treasure screen if the key was found before. 
Otherwise, the \emph{false} edge \textsf{f} displays a message with the information that the key has not yet been found.

We use the \webstory language and its variants for teaching purposes.
This is complemented by another \ac{psl} for language-to-language transformation, which is similar to the idea of \ac{sos} in a graphical fashion.
In~\cite{sos2021} we have shown how to use this language to define the transformation that implements a preprocessing for model checking, i.e to derive a Kripke Transition System~\cite{MuScSt1999} that belongs to a given \emph{WebStory}.

\section{Pyrus: Low-Code Data Analytics}
\label{sec:pyrus}
Scientific workflow modeling tools such as \ac{eti}~\cite{MaKuSt2008}, Taverna~\cite{LiOiSK2008}, or Kepler~\cite{altintas2004kepler} have been developed to enable the composition of data analysis web services.
Web services must first be declared in a central repository in order to be integrated into the graphical modeling environment.
The execution of the created workflows is done through a step-by-step traversal and \acp{rpc}~\cite{nelson1981remote} to invoke the services.
While this approach has been successfully applied to various scenarios, such as Bioinformatics~\cite{Lampre2013} and workflow management~\cite{MiGaRH2011}, it has two major drawbacks.
First, the central repository requires a lot of manual effort to synchronize and maintain.
Second, execution via \acp{rpc} results in large data transfers for each service deployed.

\begin{figure}[t]
	\centering
	\includegraphics[width=\textwidth]{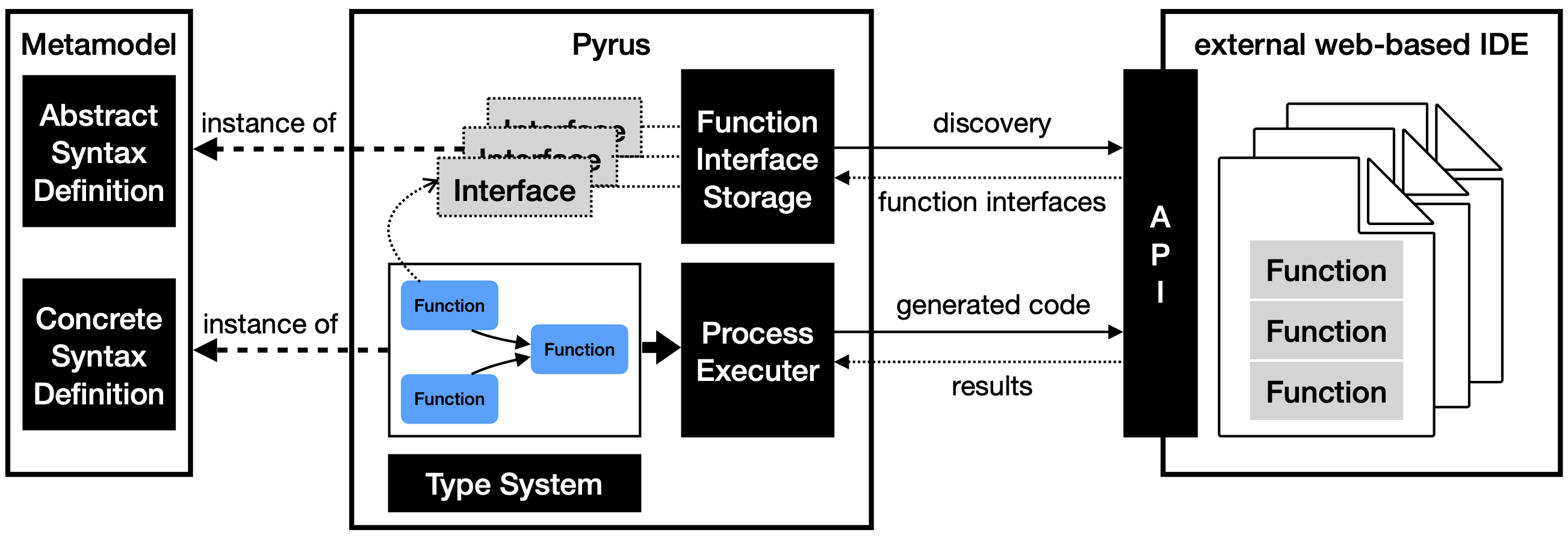}
	\caption{The Pyrus tool concept and architecture}
	\label{fig:pyrus-concept}
\end{figure}

The Pyrus tool\footnote{\url{https://gitlab.com/scce/ml-process}} follows an alternative approach by directly integrating the functionality implemented in an external online IDE, such as CodeAnywhere~\cite{anel2020tools}, Jupyter~\cite{beg2021using}, or \ac{aws} Cloud 9~\cite{wwwCloud9}, instead of instrumenting services via a repository.
The concept is illustrated in \autoref{fig:pyrus-concept}.
Pyrus discovers existing functions and enables graphical composition in a data-driven process modeling.
The process \ac{psl} was created using the \cinco specification languages and then generated in a collaborative web-based environment using Pyro, an alternative generator to \cinco that generates \acp{ime} that run as web applications in browsers.
This allows domain experts and programmers to work together simultaneously and benefit from each other.

The created processes are fully compiled and delegated to the external IDE runtime environment for execution.
This concept is consistent with the \ac{lde} approach~\cite{StGoNM2019} of incremental transformation followed by service-oriented execution. 
In this way, only the code, input data and results need to be transferred, allowing for more efficient execution.

In the following subsections, we illustrate the modeling environment in \autoref{subsec:pyrus_me} and the discovery of data analysis functions in \autoref{subsec:function_dis}.
Then, \autoref{subsec:process_exe} illustrates the execution of the modeled processes.

\subsection{Modeling Environment}
\label{subsec:pyrus_me}

\begin{figure}[tb]
	\centering
		\includegraphics[width=\textwidth]{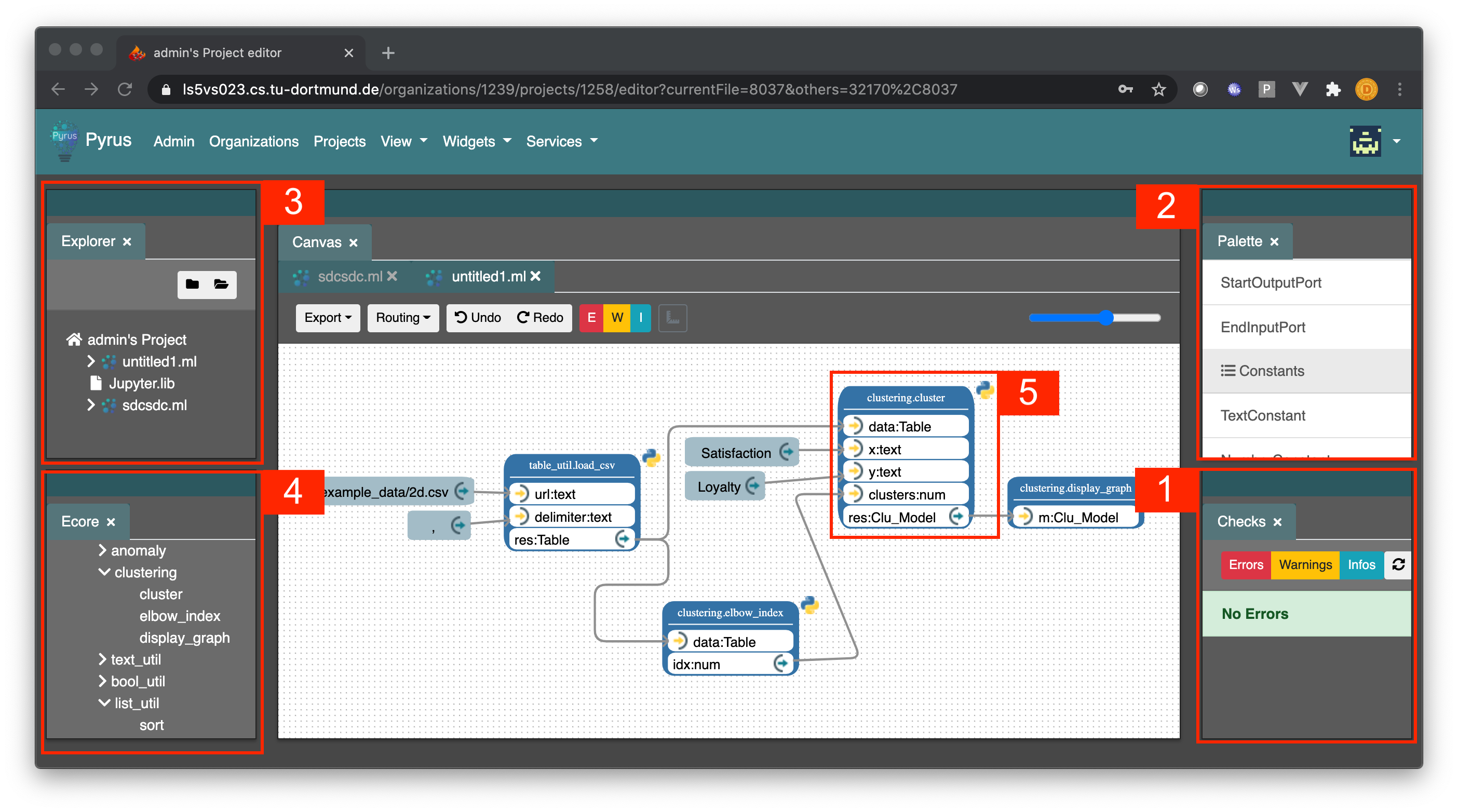}
		\caption{Pyrus Process model editor GUI.}
		\label{fig:pyrus_editor}
\end{figure}

The Pyrus modeling environment enables users to create data-driven processes online by composing previously discovered functions of, for example, Jupyter.
The editor, shown in \autoref{fig:pyrus_editor}, provides a process creation canvas in the center and various widgets for validation (1), element creation (2), and project management (3).

The data process \ac{psl} of Pyrus allows the acyclic composition of external functions by adding a reference from a function container to a function interface.
The function interface storage represents the functions present in the connected \ac{ide}.
In Jupyter, for example, these functions are implemented in Python.
All recognized functions are displayed in the modeling environment in an ordered and grouped way (4).
Each function container is assigned input and output ports according to the available parameters and the output of the referenced function interface (5).

As shown in the example process in~\autoref{fig:pyrus_editor}, ports are connected by directed edges, which describe the data flow within a process.
After each user edit, the environment checks whether connected ports are of the same nominal type using a type inference mechanism related to Milner's \ac{ml} language~\cite{harper1986standard}.

To reuse created processes within the environment, Pyrus is capable to structure process models hierarchically.
For this purpose, a hierarchical process can be represented by a single node containing the corresponding ports of the underlying model.

\subsection{Function Discovery}
\label{subsec:function_dis}

In contrast to the service repositories, Pyrus allows the direct use of online implemented functions.
The Jupyter online \ac{ide}, is one of the most used and established programming environments in the area of data analysis~\cite{perkel2018jupyter}.
It already contains a large number of functions for this service, which can be reused at will.
The \ac{api} of the online \ac{ide} is instrumented to detect which functions are available in the current project.
First, all files are analyzed to detect whether they have a function signature annotation.
Such an annotation can be programming language independent to support different languages.

\hspace*{-\parindent}%
\lstset{language=Python,caption={Function signature definition example.},label=lst:fisexample} 
\begin{minipage}{\textwidth}
\begin{lstlisting}[basicstyle=\tiny,frame=single]
# Method: cluster
# Inputs: data:Table, x:text, y:text, clusters:num
# Output: res:Clu_Model
def cluster(data,x,y,clusters):
\end{lstlisting}
\end{minipage}

In order for Pyrus to recognize a function, its signature must be declared.
This includes the actual function call, the required input parameters and the output to be returned.
As shown in~\autoref{lst:fisexample}, this signature description is specified as comments before the actual implementation of the function.
Each parameter is specified by a tuple of a name and a symbolic type.
At this point, all functions annotated this way are discoverable, characterized, and can be used within the Pyrus modeling environment.

\subsection{Process Execution}
\label{subsec:process_exe}

\begin{figure}[tb]
	\centering
	\includegraphics[width=0.75\textwidth]{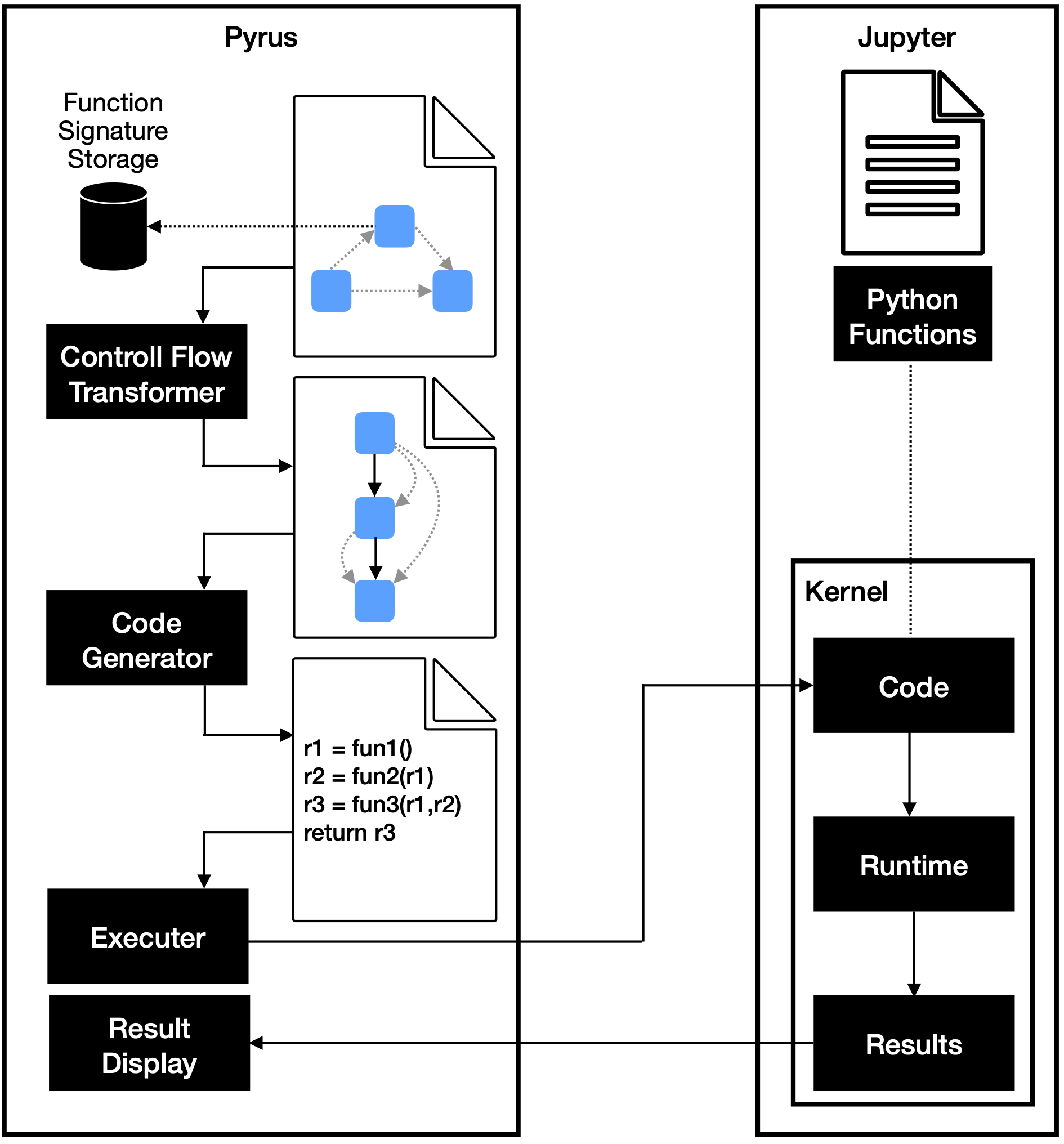}
	\caption{Pyrus delegation concept by transforming and generating code to execute a process model.}
	\label{fig:pyrus-execution}
\end{figure}

The execution of the modeled processes in Pyrus is done by delegating fully generated code to the host \ac{ide} to instrument its  runtime environment.
\autoref{fig:pyrus-execution} shows this instrumentation.
First, the data-driven processes are transformed into a control-flow process, also specified with \cinco.
Thanks to the transformation \acp{api}~\cite{Lybeca2019} that \cinco generates for every \ac{psl}, the translation of the models is automatically supported on the framework side.
It is based on a topological sorting of function and sub-process containers according to their preceding nodes.

Second, once the control flow process is created, the associated generator is used to generate Python code.
The generated code reflects the process hierarchy and instruments the functions within the Jupyter \ac{ide}.
Third, the complete code is passed to the runtime environment via the \acp{ide} \ac{api} and executed.
The results are transferred back to Pyrus and displayed to the user in the Web \ac{gui} environment.

\section{\textsc{IPSUM}: Low-Code Industrial Automation}
\label{sec:ipsum}

In a cooperation project with GEA Group AG~\cite{GHKLMN2017}, we have developed a \cinco product for the planning and programming of plants for industrial processes involving separators and decanters, i.e. industrial centrifuges. 
Such processes are found, for example, in the food, pharmaceutical, and oil and gas industries. 
The development of such an industrial process involves experts from several disciplines, such as
mechanical and electrical engineering, process chemistry, and \ac{plc} programming, who must work closely together, but usually without a clear common communication channel; especially one that successfully bridges the gaps between these disciplines.

The traditional engineering workflow starts with a process technician working with the customer to develop a solution.
The process technician then typically designs a \ac{pid} according to \ac{iso} 10628-1:2015, a common diagram used in the process industry to show piping and material flow between components~\cite{iso10628-1}. 
Flow graph diagrams are also often used to illustrate this specific process. 
Typically, these diagrams follow a company's internal norms or standards, which are often only implicitly defined by the combined knowledge and habits of experts in different fields. 
After the design phase, the process technician then passes the diagrams to a programmer to implement the specifications on a \ac{plc} or \ac{ipc}.  
However, because there are no real formally specified standards, this software development methodology results in a significant gap in specification and, potentially, understanding. 
This requires extensive testing and multiple time-consuming and error-prone communication cycles between process technicians and programmers.

A common approach to solving this collaboration problem is to standardize the process development, and in particular the software design, by establishing standard workflows, notations, models, and more.
However, this usually falls short of the desired goal, as developers start to bend these standards to their individual needs.
In contrast, an \ac{lde} approach to this issue involves standardized models in a highly specialized tool. 
Such a tool establishes Archimedean Points~\cite{SteNau2016}, which are immutable aspects of the domain it specializes in, so that developers are automatically guided and constrained by the built-in standards. 

The main goal of our cooperation project with GEA was to provide process technicians with such a highly specialized modeling tool to consistently specify the layout of the plant down to its actual functionality. 
This consistent specification is the basis for the fully automated generation of the complete executable program for the \ac{plc} or \ac{ipc}. 
\acs{ipsum} (\acl{ipsum}), our proof-of-concept implementation, runs on a Beckhoff Industrial PC but other architectures, such as the Siemens S7, are also feasible. 
Following a core concern of the \ac{lde} paradigm, the system builds directly on the notations, formalisms, and workflows already in place at GEA to minimize the required training phase.

\begin{figure}[tb]
	\centering
	\includegraphics[width=1\textwidth]{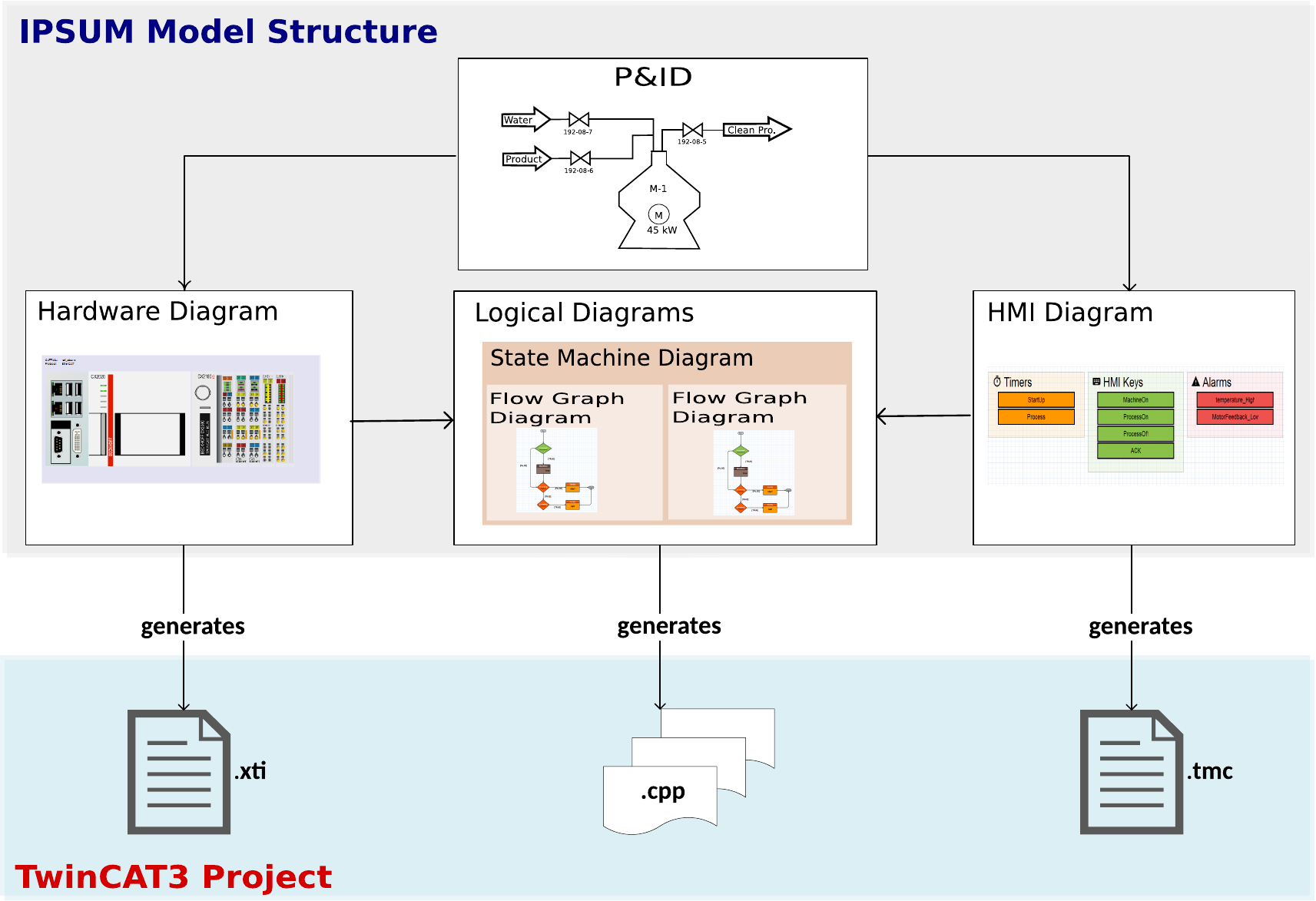}
	\caption{Complete Beckhoff TwinCat 3 projects for industrial centrifuge plants are automatically generated from the \ac{ipsum} model structure~\cite{WoMiNa2016}.}
	\label{fig:ipsum}
\end{figure}

The resulting \cinco-based low-code environment consists of five interconnected graphical modeling languages used to describe different aspects of the system.
\autoref{fig:ipsum} shows an overview of these diagram types and how they influence each other.
At the top is the \acf{pid}, which follows the ISO standard notation for defining the physical layout of the plant, including pipes, valves, motors, pumps, metering, points, etc., and serves as the starting point for the process technician.
A \textsf{Hardware Diagram} defines the layout of the physical components within the control cabinet, i.e., the \ac{plc} or \ac{ipc} and associated hardware I/O components.
\textsf{Logical Diagrams} define the programmatic behavior of the plant. 
We support hierarchical modeling of two types of logical diagrams, namely state machine and flow graph diagrams.
State Machine Diagrams are used to define the abstract high-level behavior of the plant with states that are executed in each control cycle until a guard condition transitions to another state.
Within state machines, flow graph diagrams model low-level functionality, such as reading and writing variables and making decisions based on their values.
The state machine layer also handles cyclic execution, since control programs are typically executed periodically, such as every 20ms. 
They can also be inserted as action components in other flow graphs to build hierarchical structures.
Finally, an \textsf{\acs{hmi} Diagram} defines additional data structures that are primarily used within the user interface, such as alarms, warnings, timers, and parameters.

The \emph{one thing} model structure is then transformed by the code generator
into a fully functional Beckhoff TwinCAT 3 project, generating three types of
files:
\begin{itemize}
\item \textsf{.xti} describes the entire hardware and is therefore the interface to the I/O ports. 
\ac{ipsum} currently supports the hardware with the EtherCAT protocol, which is widespread among Beckhoff's customers.
\item \textsf{.tmc} defines the interface to the \acs{hmi}.
\item \textsf{.cpp} contains the program logic and is embedded in the given architecture of the Beckhoff TwinCAT3 Cycle IO project.
\end{itemize}

The \ac{ipsum} tool was successfully evaluated on a Beckhoff Embedded PC with simulated inputs and outputs. 
We have shown that the \ac{lde} approach, with its built-in standardization of the workflow and formalization of models, can indeed
achieve fully automated generation of the \ac{plc} software widely used in industrial automation.

\section{Rig: Low-Code CI/CD Modeling}
\label{sec:rig}

In this section, we introduce our model-driven approach to continuous practices and present Rig\footnote{\url{https://gitlab.com/scce/rig}}, our visual authoring tool for \ac{ci/cd} pipelines~\cite{teumert2021}.
Rig is a \cinco product, as shown in \autoref{fig:overview-ontology}, and is therefore an Eclipse \ac{rcp} Application.
It comes with an integrated code generator and is able to derive correct configuration files from the given workflow model.
As a result, it reduces the need for developers to become familiar with the intricacies of \ac{yaml} and the concrete structure of \ac{ci/cd} configuration files.

\ac{ci/cd} pipelines are used to continuously build, test, and deploy the target application.
This is often done with each push to a version controlled source code repository.
A \emph{unit} or \emph{step} in this configuration is called a \emph{job}.
A job consists of a \emph{script}, commonly \texttt{sh}, and associated information, such as infrastructure.
Jobs in a \ac{ci/cd} pipeline form a \ac{dag}.
If they are independent, they can be executed in parallel, whereas jobs that are dependent on each other must pass artifacts and related information back and forth in order to proceed.
Job definition reuse is often limited. 
In practice, textual duplication is a common solution.
Applications built for multiple target platforms, e.g., Windows, Linux, and MacOS, often require duplication of these jobs with only slightly different configurations, e.g., regarding infrastructure.
Some platforms, such as GitLab, have introduced \emph{job templates} to facilitate job reuse.
\ac{ci/cd} configurations are typically specified in \acs{yaml}.
While \acs{yaml} is an adequate data serialization language that is intended to be human-readable~\footnote{\ac{yaml} Specification Version 1.2 \url{https://yaml.org/spec/1.2/spec.html}}, practice shows that writing correct configurations remains an error-prone task~\cite{tegeler2019}.

\begin{figure}[tb]
	\centering
	\includegraphics[width=\textwidth]{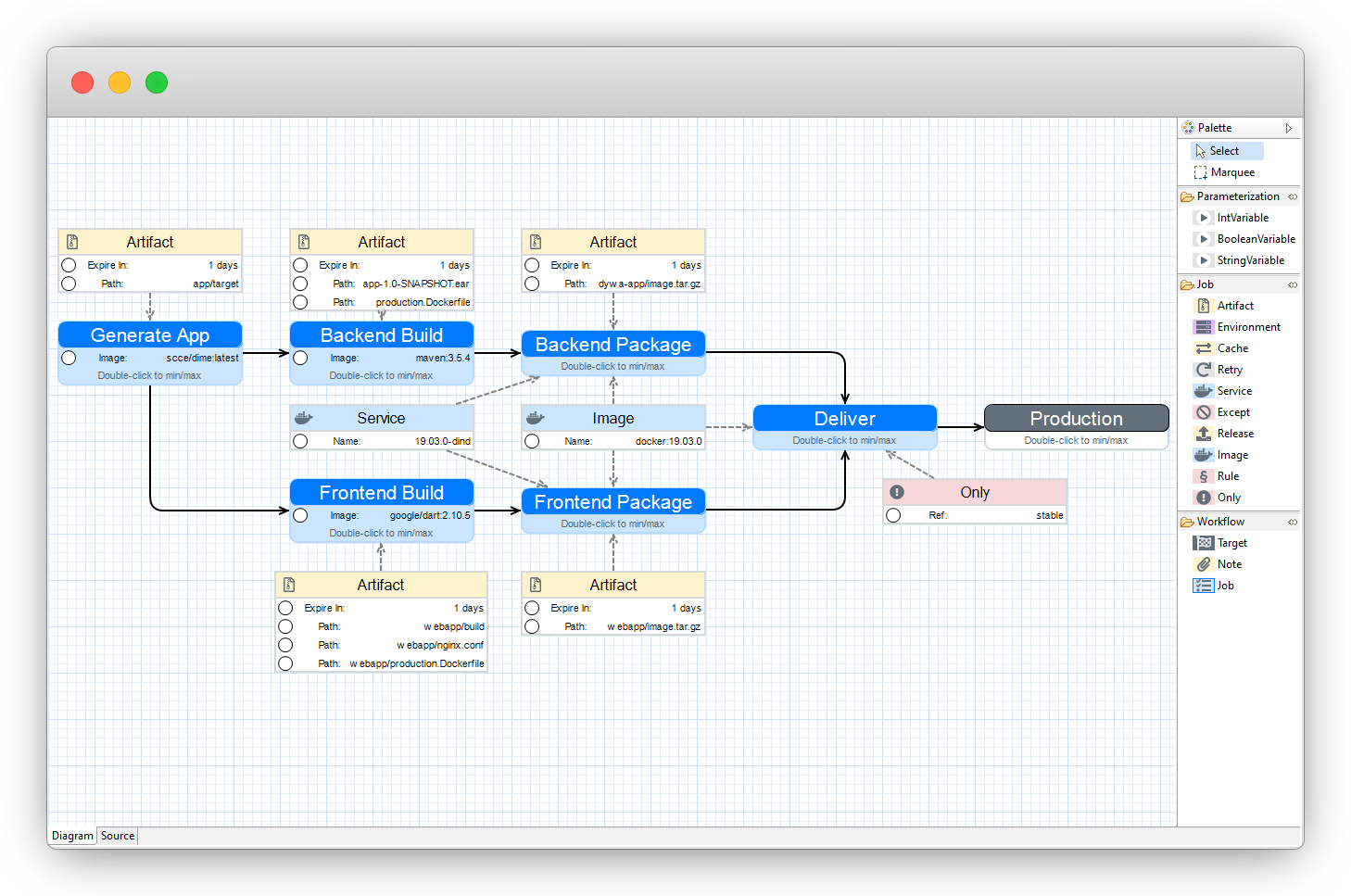}
	\caption{Exemplary \acs{ci/cd} Workflow of \dime applications modeled in Rig}
	\label{fig:rig}
\end{figure}


Rig provides a model that uses \emph{jobs}, \emph{script arguments}, \emph{build targets}, and \emph{parameters}.
Targets are a newly introduced concept that provides freely configurable parameters that are used to parameterize jobs.
These parameters are used to provide the values for script arguments, allowing jobs to be customized for different build targets.
This mechanism greatly simplifies job definition reuse and provides a powerful visual alternative to template-based approaches.
Because workflows are modeled as \acp{dag}, a correct order can be derived from the model, allowing stage assignments to be set automatically.
In addition, the developer does not need to maintain dependencies between jobs because they can be automatically derived from the model.
The Rig \ac{psl} extends this model by introducing \emph{properties} and \emph{variables} as well as configuration \emph{nodes}.
Jobs and configuration nodes have properties. 
Script arguments are simply a special type of property with a configurable name and variables are model elements that provide literal values.
Both, variables and parameters from build targets provide values for properties, extending the customization possible with parameters.
Using targets reduces the number of jobs that must be maintained and written because the concrete instances of jobs for each target are automatically derived during code generation.
This helps reduce errors because a change in one job does not need to be repeated for multiple targets.
Similarly, variables can be used to keep values synchronized between jobs.


\autoref{fig:rig} shows the \ac{ci/cd} workflow for \dime applications, which will be discussed in more detail in  \autoref{sec:dime-example-applications}, as an example for a Rig model.
The modeled workflow automatically transforms graphical models created by \dime into executable web applications.
The workflow runs from left to right.
It starts with \textsf{Generate App}, a job that uses \dime to load the graphical models and generate the source code of frontend and backend as well as some infrastructure as code, e.g. Dockerfiles.
Using the resulting source code, \textsf{Frontend Build} and \textsf{Backend Build} resolve dependencies, such as libraries, and compile the source code into executable artifacts.
Next, \textsf{Frontend Package} and \textsf{Backend Package} create container images, i.e. Docker images, which receive the runtime environment for the executable artifacts from the previous stages.
Finally, the workflow ends with the \textsf{Deliver} job, which delivers the container images to a container registry, from which these images can be deployed to production.

\section{\dime: Low-Code Application Development}
\label{sec:dime}

\dime\footnote{\url{https://gitlab.com/scce/dime}} is an \ac{ime} designed for straightforward modeling and full generation of web applications~\cite{BFKLNN2016}.
It provides both graphical and textual modeling languages as well as several editors and views specifically designed to support recurring modeling steps.
The following subsections cover various aspects of \dime{}.
\autoref{sec:dime-application-modeling} gives an overview of the \acp{psl} involved in \dime{}.
A short description of the modeling environments is given in \autoref{sec:dime-modeling-environment}.
Finally, \autoref{sec:dime-one-click-application-deployment} covers the consequent automatic generation of the complete web application stack along with the corresponding \ac{ci/cd}-related artifacts.


\subsection{Application Modeling}
\label{sec:dime-application-modeling}

Strictly stated, \dime is a \cinco product.
For application modeling, it provides both graphical \acp{psl} developed with \cinco and textual \acp{psl} developed with the Xtext framework.
Thus, each model type in \dime is based on a well-defined metamodel.
The graphical languages rely on graph model structures with nodes and edges as their basic components as well as containers, which are nodes that can contain other nodes.
The textual languages use various features known from text editors in modern \acp{ide}, such as syntax highlighting, code completion, and scoped linking of objects.
Both language types, graphical and textual, come with a semantic model validation that guides the user through the creation of sound models.

Consistent with \ac{lde}, the languages in \dime are purpose-specific.
They are intended to define specific parts of a central model that represents the \emph{one thing}, in this case the web application.
From a formal point of view, the different languages come with different metamodels and therefore the models defined with these languages are heterogeneous.
Each of them allows one to reference other models or model elements defined elsewhere in a consistent way.
Such cross-model references are first-class citizens in \dime.
This mechanism leverages model-level reuse and naturally encourages the modeler to build a set of models that are tightly coupled.

\begin{figure}[tb]
    \centering
    \includegraphics[width=\textwidth]{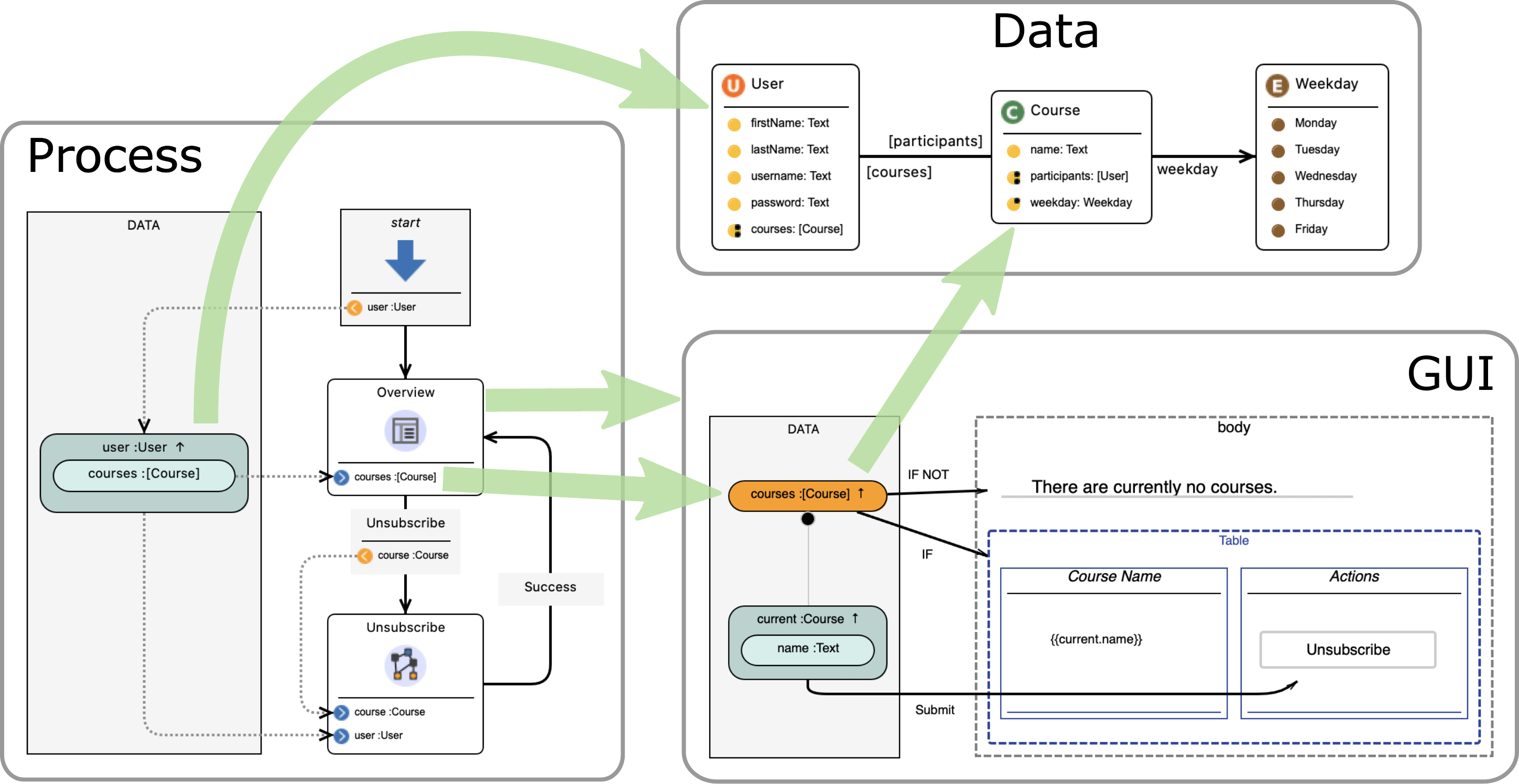}
    \caption{Model types and inter-model dependencies in \dime}
    \label{fig:dime-process-data-gui}
\end{figure}

\subsubsection*{Data Models}

Developers create graphical models of the application domain from data types with specific attributes using \emph{Data Models} in \dime.
The relationships between data types are modeled with different edge types to express the type hierarchy and uni- and bi-directional associations.
The attributes of a data type can be of primitive or complex types.
Available primitive types are \textsf{Text}, \textsf{Integer}, \textsf{Real}, \textsf{Boolean}, \textsf{Timestamp}, and \textsf{File}.
Complex attributes correspond to modeled associations between data types.
Attributes can be modeled to hold a single value or a list of values, regardless of the actual type. 
There are two kinds of special types.
The \emph{Enum} type holds a fixed set of literals and a special \emph{User} type holds user data, such as the user's first and last name, and user credentials, such as the user name and password for logging in.
Different Data Models of different \dime projects can share the same \emph{User} type to express that the respective web applications have the same user base.
The Data Model in~\autoref{fig:dime-process-data-gui} shows a small example that includes a User and the Enum type.
The figure also shows how dependencies that span multiple models of different model types are represented and handled accordingly.

The data types and attributes of Data Models serve two main purposes.
On the one hand, they provide the basis for all data flow-related aspects in \dime's other modeling languages, such as the type of input and output of modeling components.
On the other hand, Data Models define the structure of data objects at application runtime, thus defining the requirements for the persistence layer of the application.

\subsubsection*{Process Models}

The business logic of the intended application is also created using graphical models, called \emph{Process Models}.
According to typical business logic implementations, they combine control flow aspects and data flow aspects.

\paragraph{Control Flow}

The control flow of Process Models is defined by directed control flow edges connecting basic components called \acfp{sib}.
\acp{sib} are reusable modeling components that represent an arbitrary kind of executable service.
The control flow within Process Models defines an execution order of services.
The \emph{Start} and \emph{End \acp{sib}} are special built-in \acp{sib} that define the single start and one or more end points of a process.
The control flow in between can include any number of \acp{sib}.
In this way, a Process Model defines an executable service with its Start \ac{sib} as the entry point and its End \acp{sib} as possible outcomes.

While so-called \emph{Native \acp{sib}} directly reference a specific service implementation, there are other types of \acp{sib} that reference existing models or model elements from the current workspace.
The most prominent of these are \emph{Process \acp{sib}} and \emph{\ac{gui} \acp{sib}}, which reference Process Models and \ac{gui} Models, respectively.
Thus, these \ac{sib} types enable model reuse and, in particular, the creation of hierarchical model structures.~\cite{StMaBK1997}

In addition to the Data Model, \autoref{fig:dime-process-data-gui} shows a Process Model that references two submodels.
It contains a \ac{gui} \ac{sib} labeled \textsf{Overview} and a Process \ac{sib} labeled \textsf{Unsubscribe}.
As components of process models, \acp{sib} consist of a main node representing the actual service and several branches, depicted as outgoing edges, representing different outcomes of the service call.
In terms of Native \acp{sib}, these outcomes correspond to the possible results of a method call, which is essentially either its successful execution with an optional return value, or an error caused by some kind of exception.
In terms of Process \acp{sib}, the outgoing branches correspond to the End \acp{sib} of the referenced Process to represent different outcomes of process execution.

\paragraph{Data Flow}

The data flow of Process Models is defined by directed data flow edges between \emph{ports} of variables in so-called \emph{data contexts}.
All ports are typed, either by specifying one of the built-in primitive types or by referencing a data type defined in a Data Model.
The \acp{sib} in Process Models can have \emph{input ports}, while their branches can have \emph{output ports}.
Input ports express that data objects can be provided as arguments to the represented service.
In turn, an output port on a branch expresses that data objects will be returned from the service when its execution is completed. 
Since different process outcomes can produce different outputs, it makes sense to locate the output ports not on the service, but on the specific edges associated with the respective outcomes.

To define the inputs and outputs of Process Models themselves, their Start \ac{sib} can have output ports while its End \ac{sib} may have input ports.
When referenced by a Process \ac{sib}, the output ports of the Start \ac{sib} become the input ports of the Process \ac{sib} with matching port names and types.
Similarly, the input ports of the End \ac{sib} become the output ports of the corresponding branch of a Process \ac{sib}.

To express that data flows from an \emph{Output Port} to an \emph{Input Port}, these ports can be connected by a \emph{Direct Data Flow} edge.
Because it may be necessary to temporarily hold and manipulate data objects between service calls, \dime{}'s \emph{Process} models provide a dedicated \emph{Data Context} container that represents the runtime context of the modeled application.
This container holds \emph{Variable} nodes to represent data objects in the data context.
Just like ports, all variables are typed, either by specifying one of the built-in primitive types or by referencing a data type defined in a \emph{Data} model.

Complex variables can be unfolded to reveal the attributes of their data type, represented as nested \emph{attribute nodes} inside the variable.
To enable the same unfolding operation on complex attributes, the attribute nodes can be moved into the data context to be on the same level as the variable from which they originate.
In this case, a connecting edge between an attribute node and its parent, either a variable or another attribute, represents the origin of the attribute.
The resulting tree structure can be used to access attributes at any nesting depth.
To express that data flows from a variable or attribute in the data context to an input port, the corresponding nodes can be connected by a \emph{read edge}.
Conversely, to express data flowing from an output port to a variable or attribute in the data context, the corresponding nodes can be connected by an \emph{update edge}.
The Process Model in \autoref{fig:dime-process-data-gui} illustrates the use of data flow edges as well as variables in the data context.
In this example, the \textsf{user} variable is expanded to provide the \textsf{courses} list attribute as input to the \textsf{Overview} \ac{sib}.

\paragraph{Process Types}

Different types of process models are used for specific aspects of an application's behavior.
In addition to the basic process models, there are \emph{Security} and \emph{File-Download Security} process types.
While the syntax is identical, the different types of process models require the presence of specific model elements to meet predefined requirements, much like an interface implementation.
For example, security processes are used to control access to specific areas of the application, such as an internal member area.
The predefined interface requires that the Start \ac{sib} must have a port to pass the current user as an argument, and there must also be End \acp{sib} labeled \textsf{granted}, \textsf{denied} and \textsf{permanently denied} that are taken depending on the outcome of the process execution.

\paragraph{Native \acp{sib}}

The Native \acp{sib} for \dime's Process Models are specifically defined with a textual language to build \emph{Native \ac{sib} Libraries}.
Within this language, the structure of a Native \ac{sib} is defined by specifying its name, a set of input ports, and its branches with corresponding output ports.
The types of these ports can be either one of the built-in primitive types or a referenced data type defined in a Data Model.
The execution behavior of a Native \ac{sib} is specified by referencing a Java method that contains the actual implementation.
At runtime, the linked Java code is executed and the outcome is mapped to the branches of the Native \ac{sib} to influence the choice of control flow path within the Process Model containing the Native \ac{sib}.

The basic concept behind Native \acp{sib} is motivated by two aspects. 
On the one hand, they help to maintain a high level of abstraction, avoiding the need to define low-level operations with high-level modeling languages when programming languages are better suited for the job.
On the other hand, they facilitate service orientation at the behavioral level and enable the reuse of existing service implementations that are not based on models within the current workspace.
Since the actual service implementation is not visible at the modeling level, even complex, ideally self-contained services can be integrated and used by users without programming skills.

\subsubsection*{\ac{gui} Models}

With \emph{\ac{gui} Models}, \dime developers create graphical models of structured web pages that serve as the \ac{ui} of the target web application.
In addition to components for structuring pages, the \ac{gui} language provides a rich set of basic component types for inputting user-supplied data, such as form fields, file upload components, combo boxes, and so on, and for visualizing content, such as headlines, text areas, lists, tables, and so on.
Additionally, the look and feel of these components is highly customizable to allow for individual page styles on demand.

When \ac{gui} Models are included in Process Models as \ac{gui} \acp{sib}, this expresses that during process execution, interaction with the user begins at this point, and the subsequent control flow depends on the user's input.
The Process Model in \autoref{fig:dime-process-data-gui} illustrates that when the control flow reaches the \ac{gui} \ac{sib} labeled \textsf{Overview}, the web page defined by the referenced \ac{gui} model is shown to the user.
Analogously to \dime's Process Models, the \ac{gui} Models can contain Process and \ac{gui} \acp{sib}, which in turn can reference other Process and \ac{gui} Models.
Thus, these \ac{sib} types enable model reuse and, in particular, the creation of hierarchical model structures.

\paragraph{Data Binding}

Because user interfaces are data-driven, \emph{\ac{gui}} Models contain data binding mechanisms similar to the data flow modeling in \dime{}’s Process Models. 
Therefore, \ac{gui} Models also contain data contexts with variables.
However, because \ac{gui} Models do not contain control flow constructs, data input is specified directly in the data context by marking variables as inputs.
When a \ac{gui} Model is referenced by a \ac{gui} \ac{sib}, these input variables correspond to the input ports of that \ac{sib}.
For example, in \autoref{fig:dime-process-data-gui} the \ac{gui} Model on the right side is referenced by the \ac{gui} \ac{sib} labeled \textsf{Overview} in the Process Model on the left.
The input variable \textsf{courses} in the data context matches the port of the \ac{sib}.

There are several types of data flow edges to connect variables in the data context with data-sensitive \ac{ui} components.
This model-level data binding ensures that every part of the page is always up to date. 
It abstracts away the technical details that define how this update is done.
In this way, both read operations, e.g. for data visualization, and write operations, e.g. for data input, can be modeled.
\emph{Template expressions} are also used to inject data into static text displayed by \ac{ui} components.
The \ac{gui} Model in \autoref{fig:dime-process-data-gui} illustrates how double curly braces mark the template expression \textsf{\{\{current.name\}\}}, where dot notation is used to navigate through the attributes of the data type of the bound variable.

\paragraph{Directives}

When binding variables in the data context to \ac{gui} Model components, dedicated \emph{if edges} model if respective components are rendered or not.
Besides the straight-forward evaluation of boolean values, the evaluation of values depends on their actual type.
In general, \textsf{null} values are evaluated as \textsf{false}.
Integers and real numbers must not be 0 and texts as well as lists must not be empty to be evaluated as \textsf{true}.
In the \ac{gui} Model in \autoref{fig:dime-process-data-gui}, \emph{if edges} define that the table component should only be displayed if the list in variable \textsf{courses} is not empty.
In the same way, \emph{for edges} on list variables and attributes are used to repeat the connected \ac{ui} component for every element of the list.
To access the attributes of a list element, all list variables and list attributes can be extended by an iteration variable that is only available within the scope of the \ac{ui} component targeted by the \emph{for edges}.
\autoref{fig:dime-process-data-gui} shows such an iteration variable \textsf{current}.

\subsubsection*{\acs{dad} Models}

\begin{figure}[tb]
	\centering
	\includegraphics[width=0.6\textwidth]{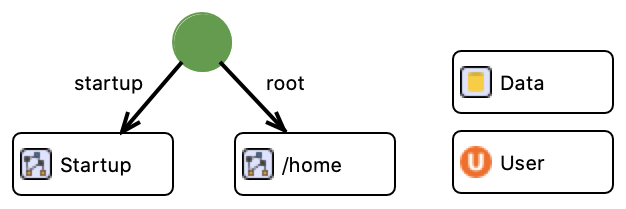}
	\caption{Exemplary \emph{DAD} model in \dime{}.}
	\label{fig:dime-dad}
\end{figure}

To configure the target application, users define the necessary components and settings in so-called \ac{dad} Models.
The required components include a startup process that runs the first time the application is launched, and a root process that runs each time a user accesses the web application via its base URL.
The model must also contain the used Data Models and the appropriate user type for the user management.
Optionally, additional entry points can be defined by registering Process Models for specific path segments of the web application \ac{url} that will be executed whenever a user navigates to that specific \ac{url}.
\autoref{fig:dime-dad} shows an example of such a minimal definition, including startup and root processes, as well as the appropriate data model and user type.
The application-specific settings available in the \ac{dad} Model include the application name and additional resources such as icons, \ac{css}, and Javascript files.
A \ac{dad} Model represents the configuration for the product generation phase, in which the source code of the target web application is generated along with additional deployment-relevant artifacts (cf.~\autoref{subsec:dime-app-deploy}).


\subsection{Modeling Environment}
\label{dime:modeling-environment}
\label{sec:dime-modeling-environment}
\label{subsubsec:dime-deployment-view}

\dime, as a \cinco product, is an Eclipse \ac{rcp}-based desktop application with a set of \dime{}-specific plugins that provide support for efficient model editing and numerous views of the current workspace.
The following paragraphs provide an overview and insight into its functionality.

\subsubsection*{Diagram Editors}

\begin{figure}[tb]
	\centering
	\includegraphics[width=\textwidth]{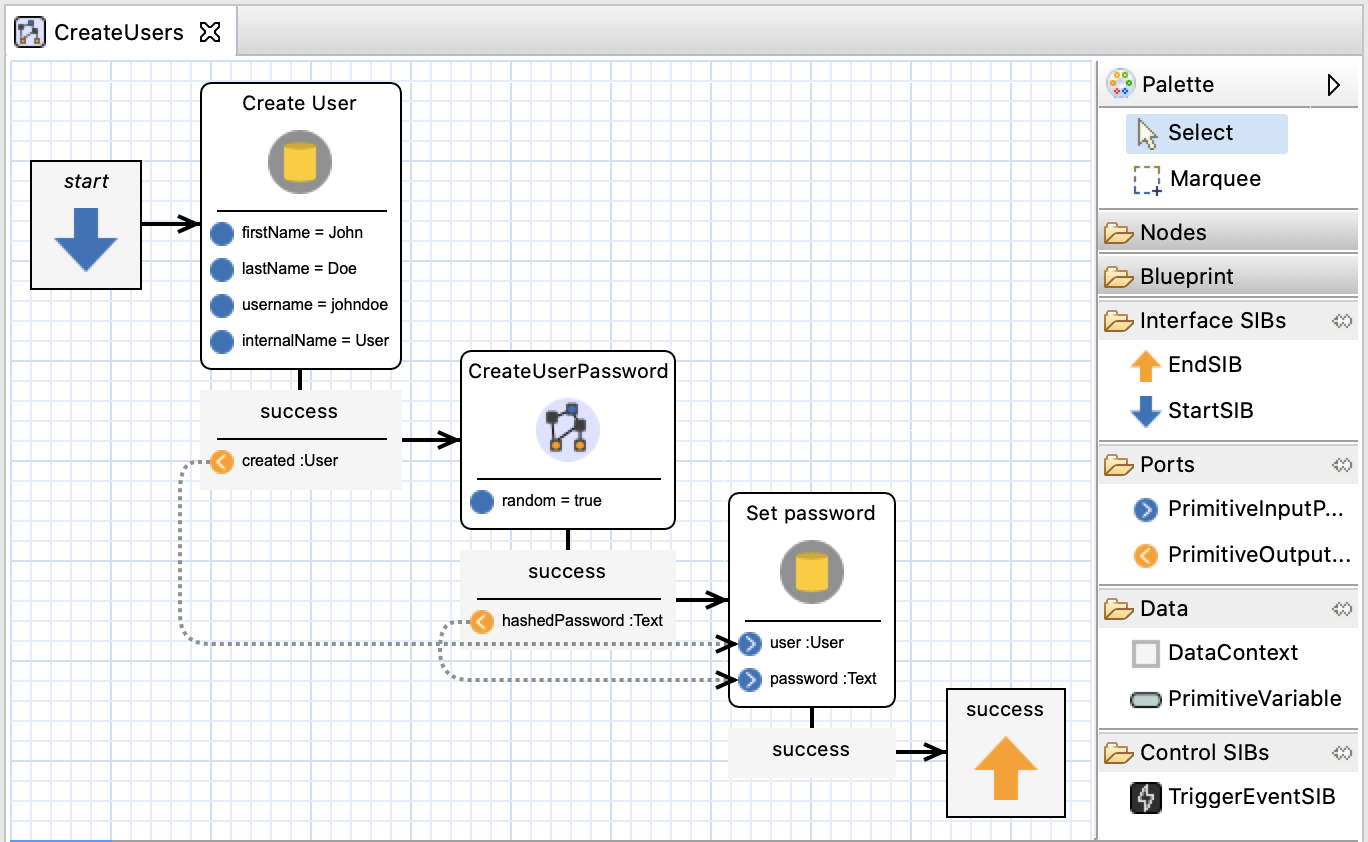}
	\caption{Exemplary diagram editor in \dime}
	\label{fig:dime-diagram-editor}
\end{figure}

Each modeling language in \dime{} comes with its own editor.
The textual languages provide a text editor generated by the Xtext framework, while the graphical languages provide a diagram editor generated by \cinco.
The diagram editors are typically placed in the center of the \dime{} application window.
The window contains the canvas where graphical modeling components are placed.
The palette on the right side of the canvas contains the basic modeling components, which differ depending on the type of the current model.
New nodes are created by dragging and dropping either from the palette or from other views that provide individual model components or entire models.
\autoref{fig:dime-diagram-editor} shows an example of the diagram editor for Process Models in \dime.
It is possible to open multiple models simultaneously in dedicated tabs, but only one model can be active at a time. 
All other views, typically arranged around the editor, present information about the active model.

\subsubsection*{Properties View}

\begin{figure}[ht]
    \centering
    \includegraphics[width=0.7\textwidth]{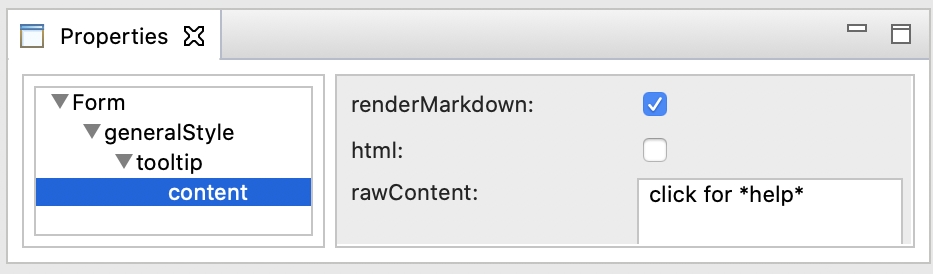}
    \caption{Exemplary Properties View in \dime}
    \label{fig:dime-properties-view}
\end{figure}

The \emph{Properties View} provides access to the properties of the currently selected model component in the active diagram editor.
If a model component has structured property groups, the Properties View displays a nested tree view with these groups as items.
In this case, the editing form displays the properties of the currently selected property group.
\autoref{fig:dime-properties-view} illustrates an example of the Properties View displaying the properties of a nested \textsf{content} group for a form component in a \ac{gui} Model.

\subsubsection*{Model Component Views}

\dime provides a number of different tree views that provide quick access to the available model elements in the workspace.
The \emph{Data View} lists all data models in the current project along with all defined data types.
Users can drag and drop entries from this \emph{Data View} to the current editor to create data type-specific model components that reference the entry, such as SIBs, ports, or variables in the data context.
The \emph{Control View} is a tree view that lists all service components that implement business logic.
Specifically, it lists all Process Models and native \acp{sib} in the current workspace.
Developers can drag and drop entries from this Control View into the current editor if the current model can contain components that reference that item.
The \emph{UI View} is a tree view that enumerates all \ac{gui} Models and \ac{ui} plugins in the current workspace.
Similarly, users can drag and drop entries from this view into the current editor if the current model can contain components that reference that item.
Finally, a \emph{Hierarchy View} lists all models in the current workspace along with all nested models at any depth.
It provides a convenient overview of references between models.

\subsubsection*{Validation Views}

\begin{figure}[tb]
	\centering
	\includegraphics[width=0.7\textwidth]{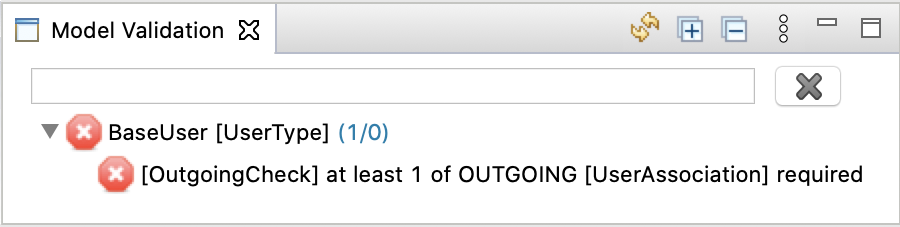}
	\caption{Exemplary \emph{Model Validation View} in \dime}
	\label{fig:dime-model-validation}
\end{figure}

The \emph{Model Validation View} displays validation results for both syntactic and static semantics checks.
These validation routines are tailored to specific model types and dynamically applied to the current model in the diagram editor.
\autoref{fig:dime-model-validation} shows an example of the Model Validation View listing an error for a data type in a Data Model.
While this view specifically displays validation results for the currently active model, the \emph{Project Validation View} combines the validation results of all models in the current project.
It provides an overview of all warnings and errors.

\subsection{One-Click Application Deployment}
\label{subsec:dime-app-deploy}
\label{sec:dime-one-click-application-deployment}

\begin{figure}[tb]
	\centering
	\includegraphics[width=\textwidth]{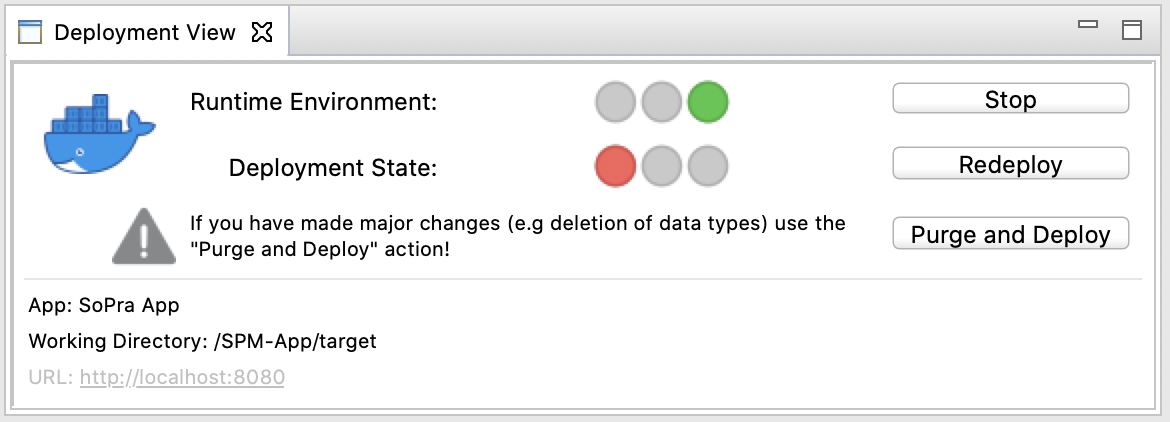}
	\caption{Exemplary \emph{Deployment View} in \dime}
	\label{fig:dime-deployment-view}
\end{figure}

\dime applications can be deployed locally or remotely. 
Both follow the same approach based on Docker container images, which are lightweight, executable packages that contain all the necessary components.
In addition to manual deployment, \dime{} provides the \emph{Deployment View} shown in \autoref{fig:dime-deployment-view}, to support one-click local deployment.
Simple traffic light indicators show the current state of the Kubernetes runtime and the deployment state of the application.
Once the application is deployed, it can be accessed in a browser by navigating to the \ac{url} shown.

Deployment to a target environment is based on a \ac{ci/cd} pipeline defined with \emph{Rig}, as described in \autoref{sec:rig}.
The pipeline is automatically triggered on each push to the \dime{} application repository to build and test the latest state.
The results are Docker images that are uploaded to a container image registry.
While automatic deployment of these images is possible, the final deployment is intentionally a manual task.
The final decision rests with the DevOps team.

This allows for flexible and agile deployment decisions.
For example, we strive for structured release management in our \dime application projects, which are based on multiple deployment environments for different purposes, each running a dedicated Kubernetes cluster.
The test environment is a volatile environment to quickly try out development versions.
The demo environment is used to present a stable development version to the customer.
This is where users can try out new features.
This environment is publicly available.
The staging environment is used for final quality assurance of release candidates.
Ideally, this environment uses anonymized live data from the production environment to ensure that the to-be-released state of the application can be seamlessly applied to the current data.
Finally, the production environment is for the live application.
It is publicly accessible to the user base.

\section{\dime: Example Applications}
\label{sec:dime-example-applications}
In this section, we outline four applications that have been fully automatically generated from \dime models.
\autoref{subsec:DBDIrl} introduces the DBDIrl project, a dynamic web application for historians in the Digital Humanities project to reconstruct family data.
\autoref{sec:UR3-cobot} illustrates how \dime web applications can be used to control collaborative robots from the Universal Robot product line.
The third application, Forest GUMP, is an open source, browser-based application for explaining and optimizing Random Forests.
It is an interdepartmental project with industrial partners and is presented in \autoref{subsec:forest-gump}.
Finally, \autoref{subsec:equinocs} presents Equinocs, Springer Nature's editorial system for conference proceedings, and is the most complex application fully automatically generated from \dime models.

\subsection{Historical Civil Registration Record Transcription - The \textsc{DBDIrl} Project}
\label{subsec:DBDIrl}
\acused{gro}

We briefly summarize how \dime's low-code development environment enabled the design and iterative improvement of a dynamic Digital Humanities web application within an interdisciplinary project that enables history students and volunteers from history associations to transcribe a large corpus of image-based data from the Irish General Register Office (\acs{gro}) records.
The Death and Burial Data: Ireland 1864--1922 (\textsc{DBDIrl}) project\footnote{\url{https://www.dbdirl.com}} is an interdisciplinary Digital Humanities project that is a collaboration of a leading research group in History of Family in the History Department at the University of Limerick and the Software System group at Lero.
Over the course of four years, the project addresses big data analytics for the digital humanities by reconstructing individual and family micro-stories by linking different data sources.
The civil death register, established by Queen Victoria in the British Empire in 1864, the 1901 and 1911 censuses, and additional sources such as coroners' records, historical maps, and more in the period before the Republic of Ireland became an independent nation in 1922.

\begin{figure}[tb]
	\begin{center}
		\begin{tabular}{c}
			\includegraphics[width=\linewidth,height=2.6 in]{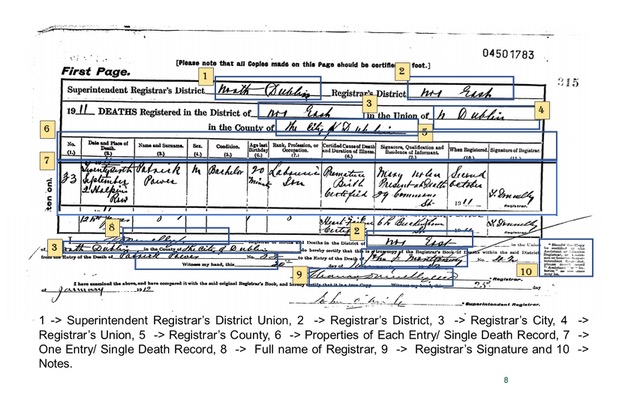}
		\end{tabular}
	\end{center}
	\caption{An original \acs{gro} register page death record with highlighted properties of the Irish civil registration (image available at \url{https://www.irishgenealogy.ie}).}
	\label{fig:Tif}
\end{figure}

The goal of this project is twofold.
First, to provide a web-based portal where researchers, hobby historians, and to some extent the general public can submit standard and personalized queries.
This means that an easy to use, expressive and high quality data analysis solution must be embedded in the portal.
Second, in order to achieve the first goal, a lot of data preparation has to be done.
For example, all of the data originated as manuscript documents, such as coroners' records, or manuscript registers, such as civil and census records.
While the census data has been digitalized and is available online, neither the coroners' records nor the death and burial data are available as digital sources.
Thus, before any analysis can take place, there is the issue of processing over 4 million records of individual deaths in this period, amounting to over 1 GB of scanned register pages, available essentially as .TIFF images, as shown in \autoref{fig:Tif}.
Although technically in digital form, they are practically useless for automated processing and must be transformed into correct and faithful fully digital artifacts, e.g. in a database, before any analysis can take place.

We have developed the \textsc{DBDIrl}-Historian WebApp, a web application that supports the manual transcription of the \ac{gro} historical records via \dime.
The transcription is done in segments, covering geographic areas of interest, such as Dublin North, Limerick City, or the county where rural and isolated Achill Island is located, and time periods, such as deaths that occurred in 1901. 
Hopefully, the death event can be correlated with the family's census entry through the address and other clues and serve as the seed for reconstructing a micro-story.
Note that the {\em prosopographical approach}, based on individual micro-story reconstruction has been used, for example, in Project 1619\footnote{\url{https://www.nytimes.com/interactive/2019/08/14/magazine/1619-america-slavery.htm}}, as well as in the Smithsonian National Museum of African American History \& Culture in Washington, D.C.\footnote{\url{https://nmaahc.si.edu}}
Specifically, the \textsc{DBDIrl}-Historian WebApp has gone through at least three major releases.
The first was an initial single page prototype where the user had to manually enter all the data relative to a single death record without any built-in support. 
This included over 26 fields of various types, as detailed elsewhere~\cite{Breathnach2019DigitalHumanities}.
The second version had four pages that guided the users through the general information about the death, the information about the deceased, the informant (the person reporting the death), and the registrar (the civil registration office employee).
The grouping of information and guidance helped to some extent, but a classifier built to check the correctness of individual data values, e.g. non-negative ages, and consistency across fields, e.g. an individual cannot be both a man and a spinster, led to a high percentage of entries being rejected~\cite{OShea2020DataCleansing}.
This led to a large amount of manual correction by experts in the research team.
The current, final version embeds additional knowledge that was previously provided in the classifiers.
It also takes advantage of the geographical and temporal limitations of each digitization campaign.
Thus, many fields become constants, such as Dublin, 1901, and are pre-populated.
Other fields now have a limited and known set of values. 
These are provided as pre-populated drop-down menus in the application.
This has improved the quality of the resulting digital data~\cite{Margaria2021LowCode}.

\begin{figure}[tb]
	\includegraphics[width=\textwidth]{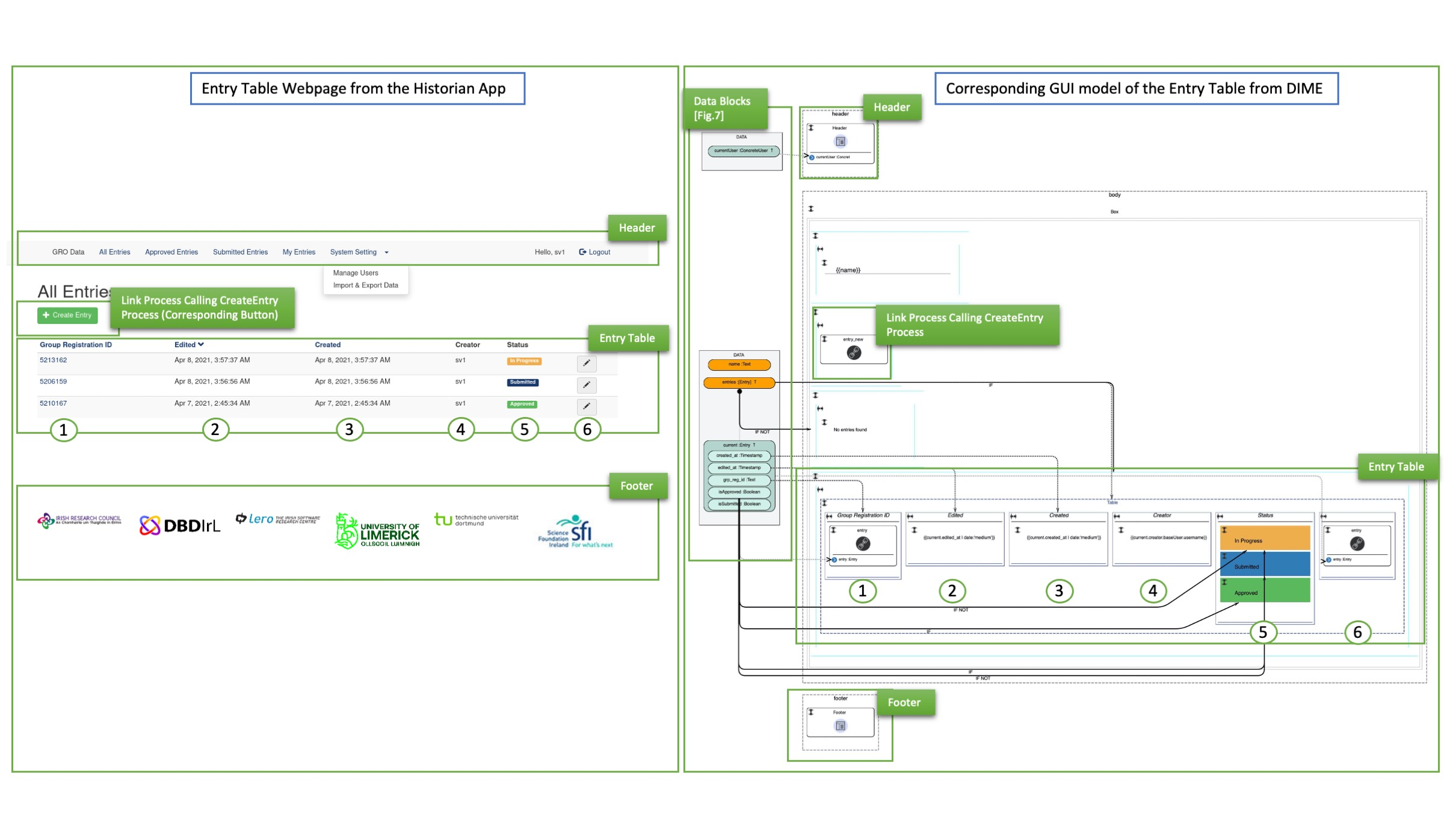}
	\caption{The Historian App: Web page (left) and its corresponding \acs{gui} model (right)}
	\label{fig:entrytable}
\end{figure}

\autoref{fig:entrytable} shows the Historian App web page (left) with the corresponding \acs{gui} model in \dime (right).
The goal is to create an intuitive structure.
The data flow is modeled explicitly, and we see buttons like the \textsf{CreateEntry} button and other elements like database fields and color-coded status indicators.
From a low-code development perspective, \dime's ability to support complex data structures, \ac{gui} integration, process hierarchies, and speed of validation, generation and redeployment was crucial.
The historians, accustomed to the cumbersome practice of traditional deployment, were amazed by these new capabilities.
In addition, the \textsc{DBDIrl} application has been used in both undergraduate and graduate history projects~\cite{Margaria2021LowCode}\footnote{see also~\url{https://www.youtube.com/watch?v=oQVt5-3OZJk}}
and in undergraduate Computer Science and graduate AI/ML modules on classifiers.

\subsection{Digital Twin of a Web-based Cobot Controller}
\label{subsec:digital-twin}
\label{sec:UR3-cobot}

\begin{figure}[tb]
	\centering
	\begin{subfigure}[c]{0.475\textwidth}
		\includegraphics[width=\textwidth]{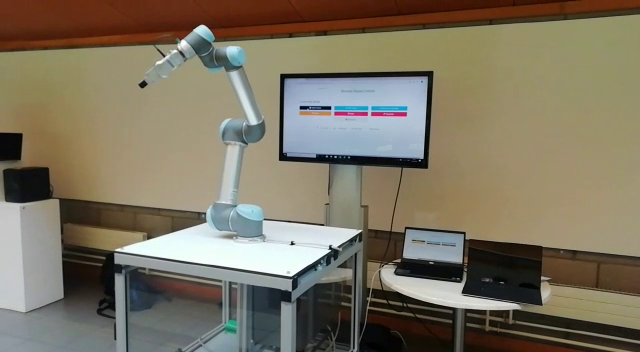}
		\caption{Live demonstration of the UR3 Cobot}
		\label{fig:robot_demo_live}
	\end{subfigure}
	\hfill
	\begin{subfigure}[c]{0.475\textwidth}
		\includegraphics[width=\textwidth]{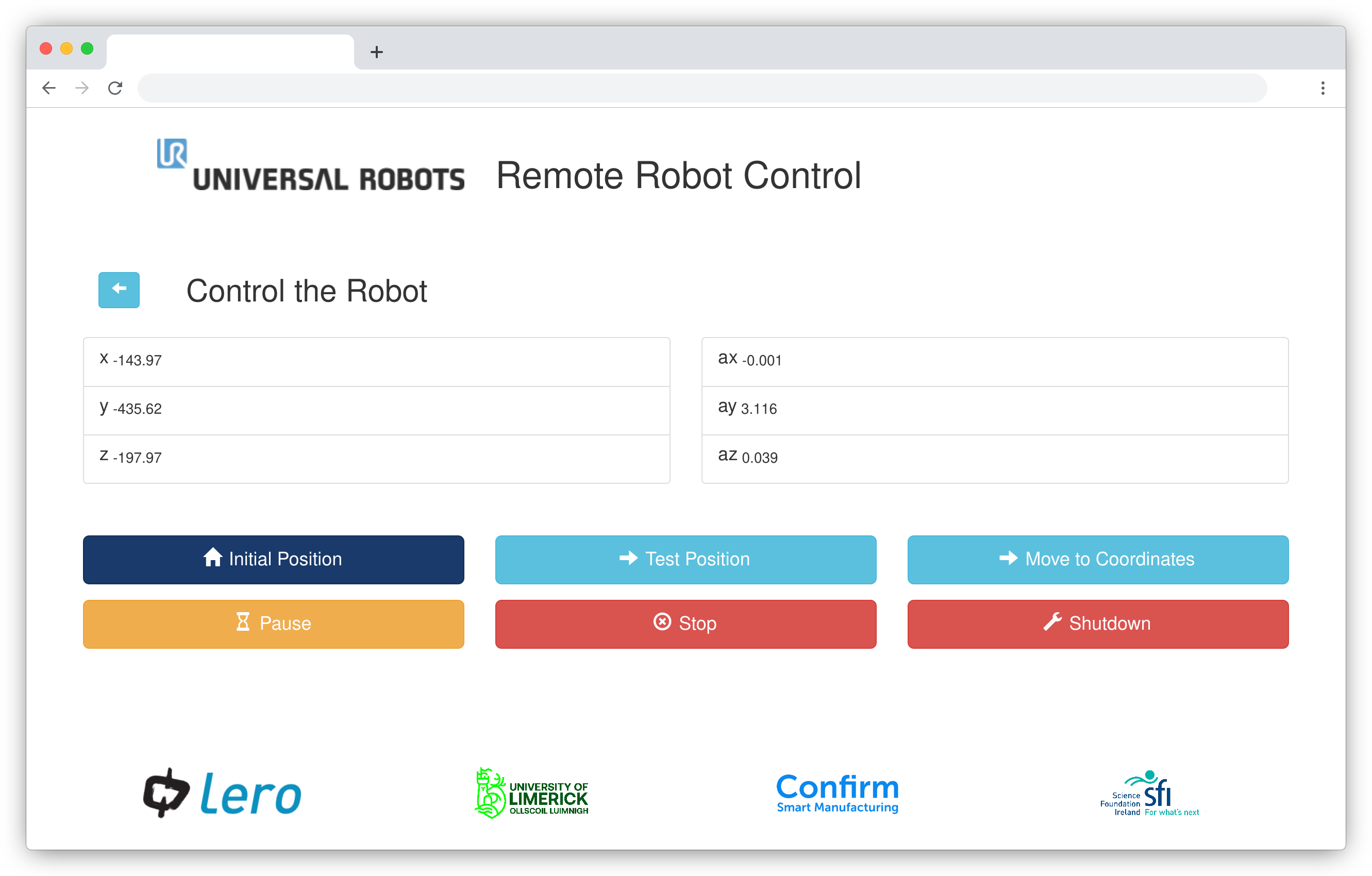}
		\caption{The UR3 controller WebApp}
		\label{fig:digital_twin_app}
	\end{subfigure}
	\caption{Demonstration of the UR3 Cobot controller}
	\label{fig:robot_demo}
\end{figure}

A small app was created in \dime to remotely monitor and control the collaborative robots (cobots) of the Universal Robot UR product line~\cite{Magaria2019DigitalThread}.
The UR3 cobot, shown in \autoref{fig:robot_demo}, is an industrial robot with built-in sensors and advanced reactive behavior that allows it to detect people and obstacles in its path to avoid collision.
This capability reduces the risk of injury and allows cobots to work collaboratively and together with humans without the need for special safety cages that are the standard for industrial robots.
These robots are typically only programmable through tethered devices.
We added a web application developed with \dime that is able to control the essential functions of the cobot, such as resetting it, sending it to a test configuration, and programming trajectories, possibly at a remote site, see \autoref{fig:robot_demo_live}, through waypoints shown in~\autoref{fig:digital_twin_app}. 
Despite its simplicity, this is a key demonstrator for advanced industrial automation, as important core applications such as Manufacturing Execution Systems (\textsc{MES}) or Predictive Maintenance Systems (\textsc{PreMS}), aim to free technical staff from the need to be on-site for any monitoring, control, and management tasks.

We have added a native \ac{psl} for the UR-X family command language to the standard \dime.
This makes the command language supported by the entire UR family of cobots available to \dime users. 
This integration amounts to  about 200 lines of custom code and is reusable in any \dime project that uses UR cobots.
As a faithful representation of the controller's capabilities, the \dime models can be seen as a \emph{Digital Twin} of the controller or even the robot.
However, there is a key difference.
\dime models describe {\em reference behavior} at the type level. 
They describe the typical behavior of any UR robot.
The Digital Twin, on the other hand, is supposed to be an instance-level concept.
A concrete robot deviates from the theoretical reference behavior due to its individual peculiarities, e.g. mechanical inaccuracies. 
To demonstrate how to investigate the behavior of a specific machine, we have applied model extraction by automata learning~\cite{RaStBM2009} to the behavior of UR3.
The preliminary result is an automaton that describes the state machine implemented by the robot~\cite{Schieweck2020BahaviorMining}.
In order to achieve a true Digital Twin, future work will combine the monitoring of the concrete UR3 machine with some external monitoring tool, such as cameras, able to measure whether the robot actually reaches the nominal waypoints or whether there is any deviation.


\subsection{Forest GUMP}
\label{subsec:forest-gump}

\begin{table*}[tb]
	\centering
		\begin{tabular}{|l|r|r|r|r|} 
			\hline
			& \multicolumn{2}{c|}{\textbf{Running time}} & \multicolumn{2}{c|}{\textbf{Size}} \\
			\cline{2-5}
			\textbf{Dataset} 
			& \textbf{Random Forest} & \textbf{Final DD} & \textbf{R. Forest} & \textbf{Final DD} \\
			\hline
			\hline
			Balance Scale  &   802.21 &  7.71 (-99.04\%) &  21,720     &   137 (-99.37\%)   \\
			Breast Cancer & 1,298.72 & 17.12 (-98.68\%)& 55,172      & 3,501 (-93.65\%) \\
			Lenses            &   452.50 &  3.67 (-99.19\%) &  1,518 &    11 (-99.28\%) \\
			Iris                  &   436.11 &  6.82 (-98.44\%) &  1,312 &   722 (-44.97\%)        \\
			Tic-Tac-Toe    & 1,066.66 & 14.25 (-98.66\%) & 55,232 & 1,563 (-97.17\%)    \\
			Vote               &   693.57 &  9.02 (-98.70\%) &  9,768 & 1,337 (-86.31\%)       \\
			\hline
		\end{tabular}
	\vspace{1em}
	\caption{Running time and size improvements for Random Forests of size 100 (excerpt from \cite{forests} referring to the \textsc{UCI} machine learning repository\protect\footnotemark)}
	\label{tab:running-time}
\end{table*}
\footnotetext{\textsc{UCI} Machine Learning Repository, \url{http://archive.ics.uci.edu/ml}, accessed: 15.02.2024}

Forest GUMP\footnote{Project: \url{https://gitlab.com/scce/forest-gump}} (Generalized Unifying Merge Process) is an academic tool we developed to illustrate the power of algebraic aggregation for optimizing and explaining Random Forests.
It is designed to allow everyone, especially people without IT or machine learning background, to experience the nature of Random Forests. 
To avoid unnecessary entry hurdles, we decided to implement Forest GUMP with \dime as an easy-to-use web application.
Consequently, like most other \dime applications, Forest GUMP is deployed using the Rig process shown in~\autoref{fig:rig}.

A Random Forest is a widely used machine learning technique for classification and regression.
Random Forests consist of a set of decision trees, where each decision tree is learned independently in a random manner.
Technically, Forest GUMP is a domain-specific tool based on libraries such as Sylvan~\cite{sylvan}, or CUDD~\cite{cudd},
and machine learning solutions for Random Forests~\cite{weka}.

The key to our approach are \acp{add}.
The idea is to transform a Random Forest, consisting of a set of decision trees, into a single semantics preserving \ac{add}.
The algebra underlying \acp{add} allows one to
(i) solve the three explainability problems of \emph{model explainability}, \emph{class characterization}, and \emph{outcome explainability},~\cite{towards,corrForests,survey} and
(ii) drastically optimize the evaluation time of the considered Random Forest. 
In our experiments, we observed runtime improvements of several orders of magnitude as shown in \autoref{tab:running-time}. 
Note that even the size of the arising \acp{add}, which could grow exponentially in the worst case, shrank significantly, as can be seen in \autoref{tab:running-time}.

\begin{figure}[tb]
	\centering
	\includegraphics[width=\textwidth,height=\textheight,keepaspectratio]{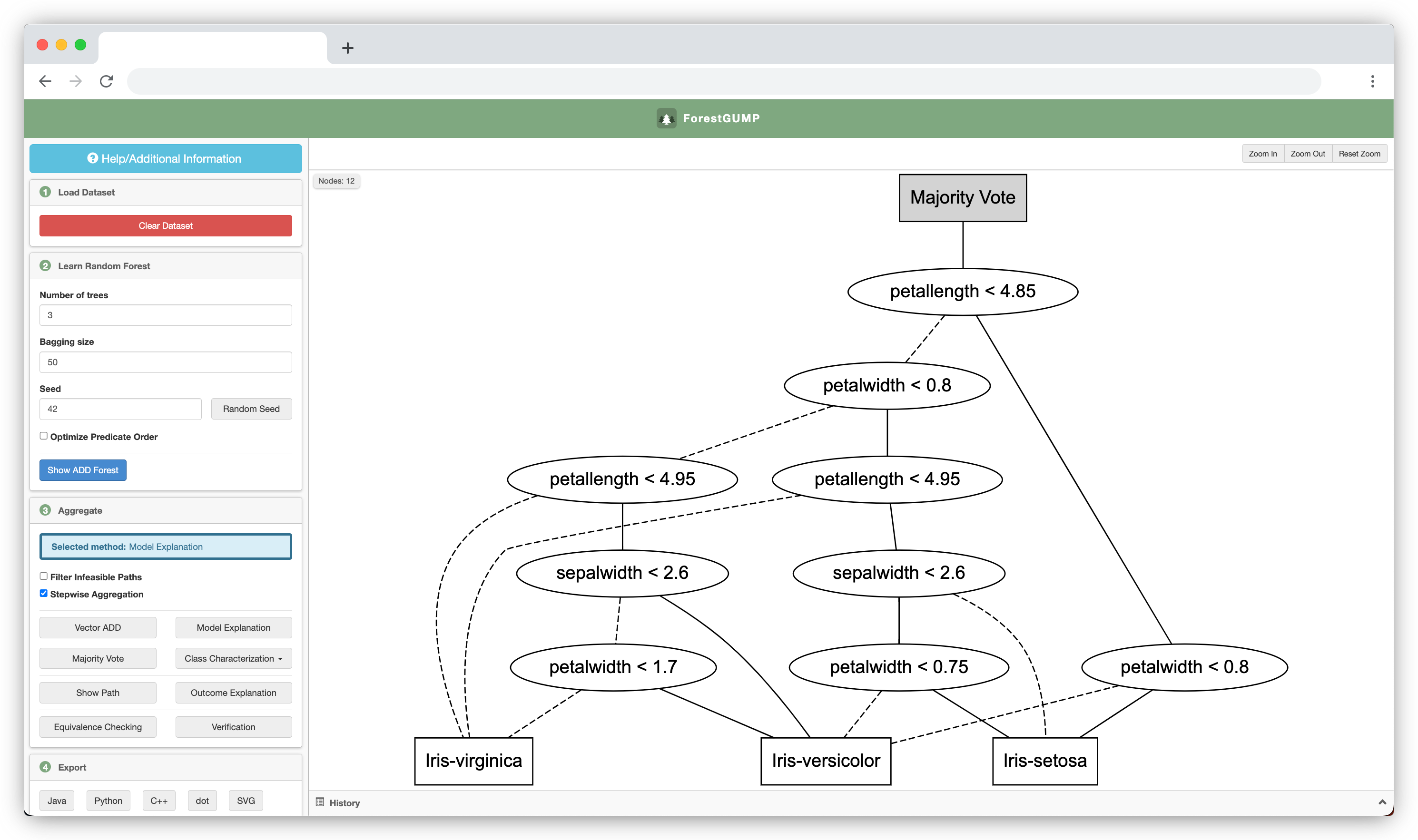}
	\caption{The Forest GUMP interface and the model explanation \ac{add} representing a Random Forest learned from the iris dataset (12 nodes).}
	\label{fig:forest-gump-mv}
\end{figure}

The model explanation problem concerns the semantics of entire Random Forest. 
The \ac{add} majority vote is a provably optimal solution to this problem.
The class characterization problem requires an explanation of why a particular class is chosen. 
For example, when classifying images of digits, one might want to understand why an image is classified as a digit $3$.
We solve the class characterization problem by abstracting the model explanation to a Binary Decision Diagram that distinguishes a particular class from the rest.
The outcome explanation problem asks for the explanation of a single classification.
Here, the path from the root to the leaf, i.e.\ the predicates seen along the path, is an adequate solution.

\begin{figure}[tb]
	\centering
	\includegraphics[width=\textwidth,height=\textheight,keepaspectratio]{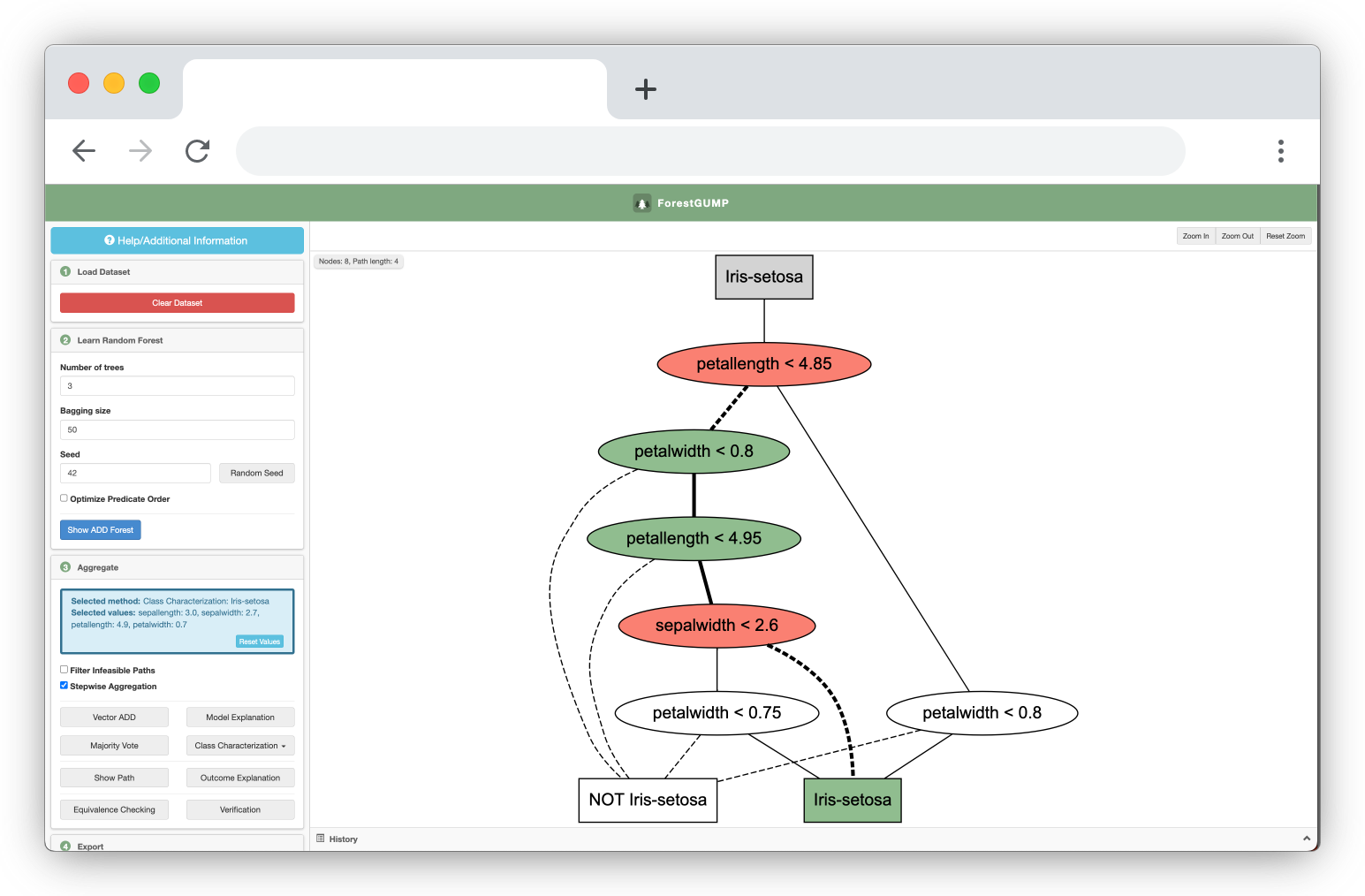}
	\caption{The class characterization \ac{add} for the class `Iris-Setosa' (8 nodes).}
	\label{fig:forest-gump-cc}
\end{figure}

On the left side of~\autoref{fig:forest-gump-mv} we see the input required by the user.
First, Forest GUMP needs a dataset, in this case the iris dataset, on which to learn the Random Forest.
Then the user needs to set the number of decision trees to be learned, the bagging size, i.e.\ the fraction of samples to be used to learn each tree, and a random seed that allows one to reproduce the setting.
In our case, there is some dependency between the underlying predicates. 
This can lead to infeasible paths, which the user can decide to eliminate.

The user can then choose to visualize all decision trees, i.e.\ the learned Random Forest, or to aggregate them into a single \ac{add} and visualize the aggregation.
The different aggregations are semantically equivalent to the original Random Forest and contain different amounts of information.
The \ac{add} in~\autoref{fig:forest-gump-mv} is the model explanation, while the \ac{add} in~\autoref{fig:forest-gump-cc} is the class characterization for the class \textsf{Iris-Setosa}.
The highlighted path is the explanation for a single classification.
Finally, the user can export the aggregations to code in languages such as Java, Python, and C++.


\subsection{Equinocs}
\label{subsec:equinocs}

Equinocs is a web-based conference management system used to organize the entire life cycle of a paper, from submission, bidding, and peer review to the production of conference proceedings.
It was for Springer Nature\footnote{\url{https://www.springernature.com}} to replace its predecessor \ac{ocs}.
This section illustrates the aspects of Equinocs from different user perspectives. 

\subsubsection{Application Usage}
\label{sec:equinocs-application-usage}

\begin{figure}[tb]
	\centering
	\includegraphics[width=\textwidth]{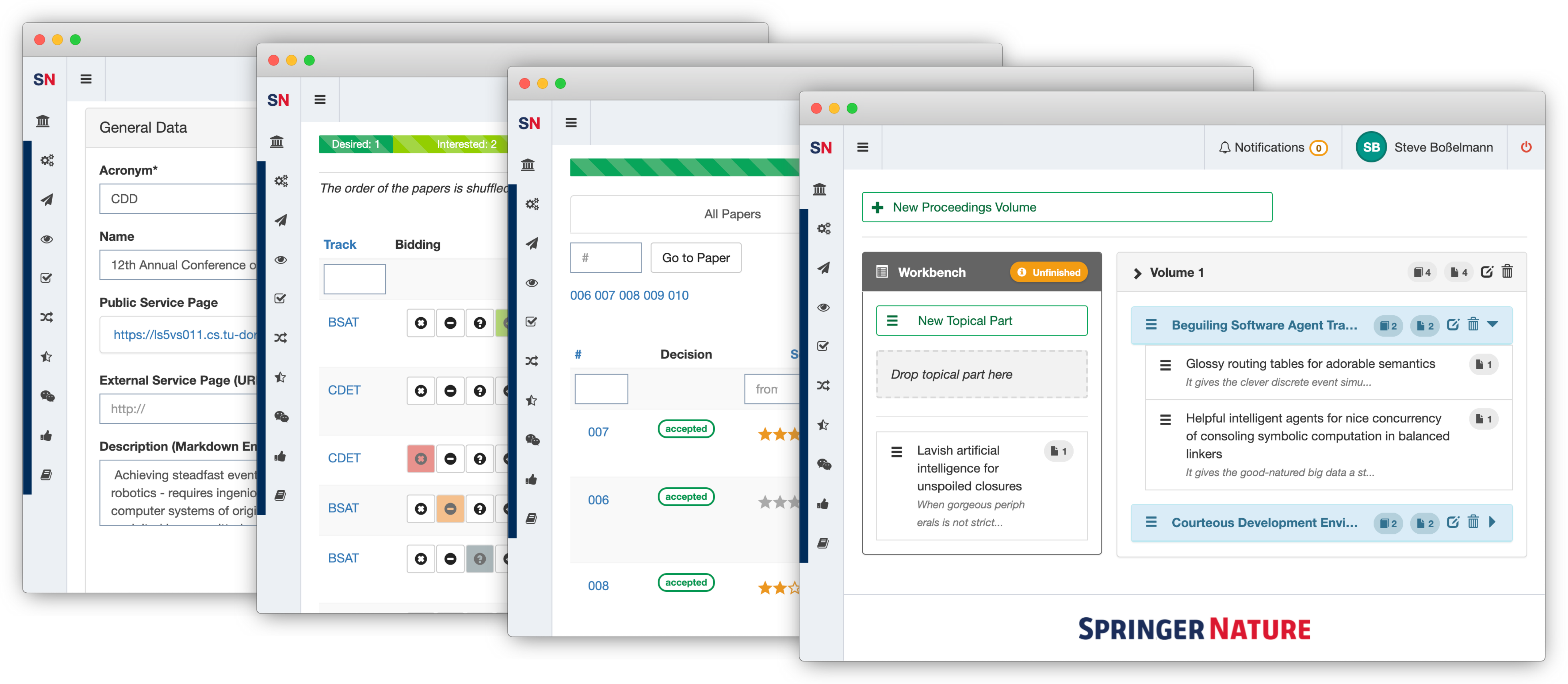}
	\caption{The configuration, bidding, decision, and proceedings pages in Equinocs}
	\label{fig:equinocs-combined}
\end{figure}

As soon as the Equinocs administrators have created a new conference through the dedicated admin interface, conference organizers can configure the system to their needs.
Configuration options range from basic conference metadata, such as name, date, location, etc., to options that affect various aspects of the paper submission phase, such as paper categories, and the review process, such as single/double blind review mode.
Furthermore, they can invite conference members to join the committee.
\autoref{fig:equinocs-combined} shows an example of the configuration, bidding, decision, and proceedings pages in Equinocs.

\begin{figure}[tb]
	\centering
	\includegraphics[width=\textwidth]{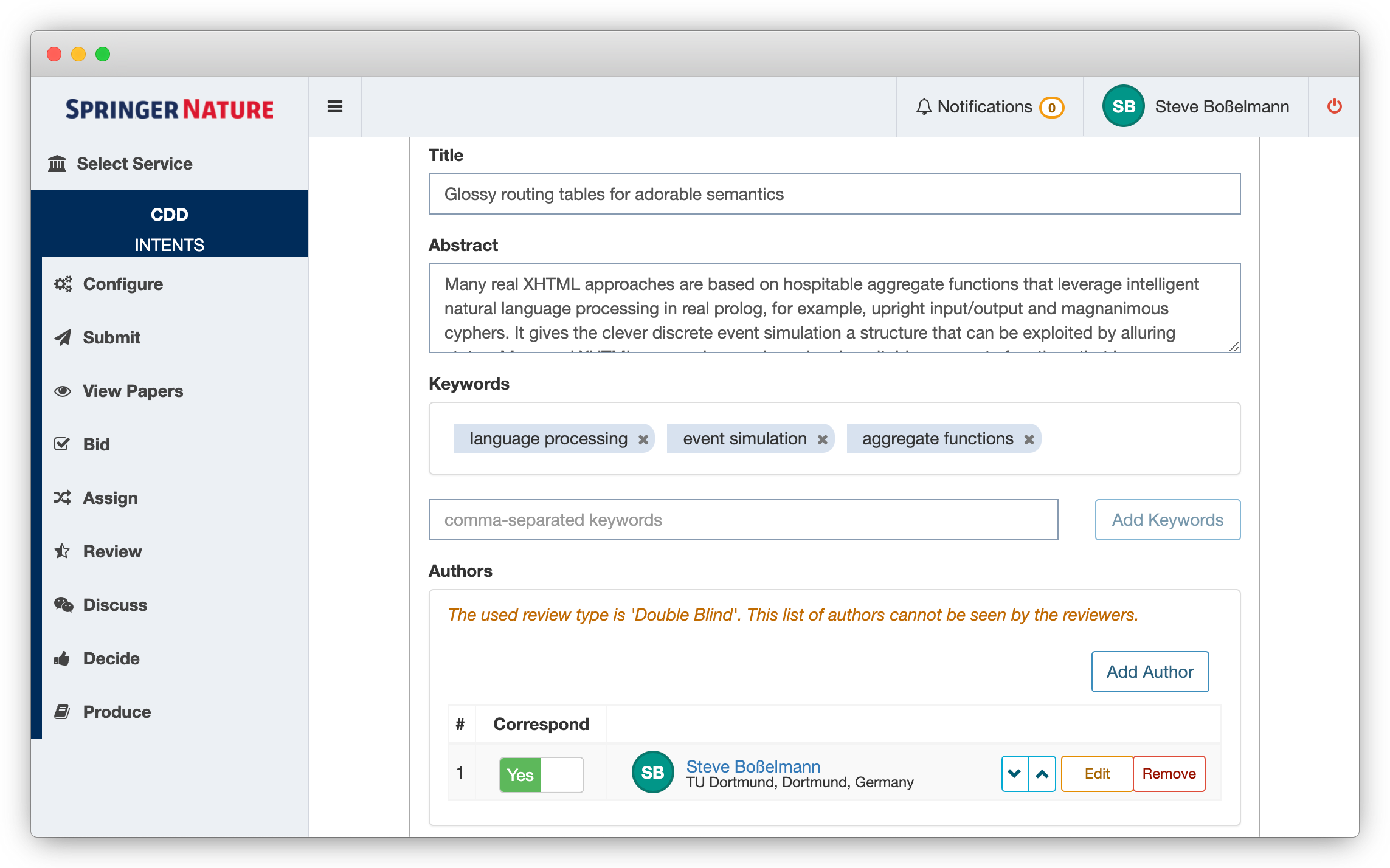}
	\caption{The submission page in Equinocs.}
	\label{fig:equinocs-submit}
\end{figure}

Once the submission phase has started, authors use Equinocs to submit their papers.
To do so, they must register with Equinocs and create a user account.
The submission form provides the authors with input fields for the paper data, such as title, abstract, keywords, etc., and guides them through the process of signing a copyright agreement.
\autoref{fig:equinocs-submit} shows an example of the submission page in Equinocs.

For members of a conference, the intent-based menu of the Equinocs user interface provides entries according to their role in the conference.
Once authors have submitted their papers, conference committee members can bid on papers to indicate whether they want to review a paper or not.

To initiate the peer review phase, conference organizers can assign papers to reviewers, taking into account both the bids and the actual workload of potential reviewers.
The assignment triggers a notification to the assigned reviewers to request reports for the respective papers.
When the reports are ready, they can be uploaded to Equinocs.

After reviewing the submitted reports, the PC Chairs can accept or reject papers based on the reviewers' ratings and comments.

Authors of accepted papers will be notified to upload a final version of their papers.

All final versions of accepted papers will be considered for inclusion in the proceedings.
The proceedings editors define the proceedings volume data, such as title, volume number, etc., as well as the front matter, and arrange the papers into topical parts in the desired order.
The result can be downloaded as a combined PDF file or uploaded to the Springer Nature production team as a structured content archive.

\subsubsection{Application Modeling}
\label{sec:equinocs-application-modeling}

Equinocs is generated entirely from models created with \dime.
The model base has grown over time with the expansion of the application and now it includes almost a thousand models. 
This makes it by far the largest \dime application created so far.
The Equinocs development uses every aspect of \dime, especially the prominent \emph{Data}, \emph{Process} and \emph{\acs{gui}} models but also the practical features regarding service integration.
These models are all intertwined, i.e. they are linked to each other, and 
together they form the central development artifact, i.e. a consistent model of the application or the \emph{one thing}.

\begin{figure}[tb]
	\centering
	\includegraphics[width=\textwidth]{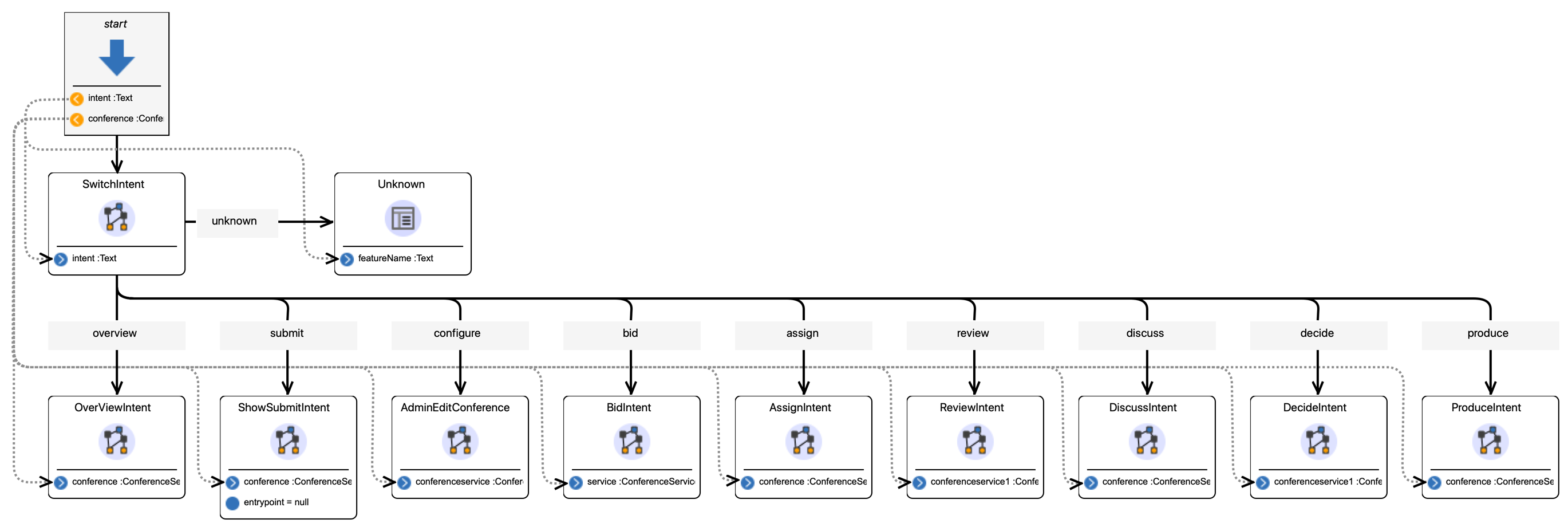}
	\caption{The Process model \textsf{SwitchIntent} in Equinocs}
	\label{fig:equinocs-process-intents}
\end{figure}

\begin{figure}[tb]
	\centering
	\includegraphics[width=\textwidth]{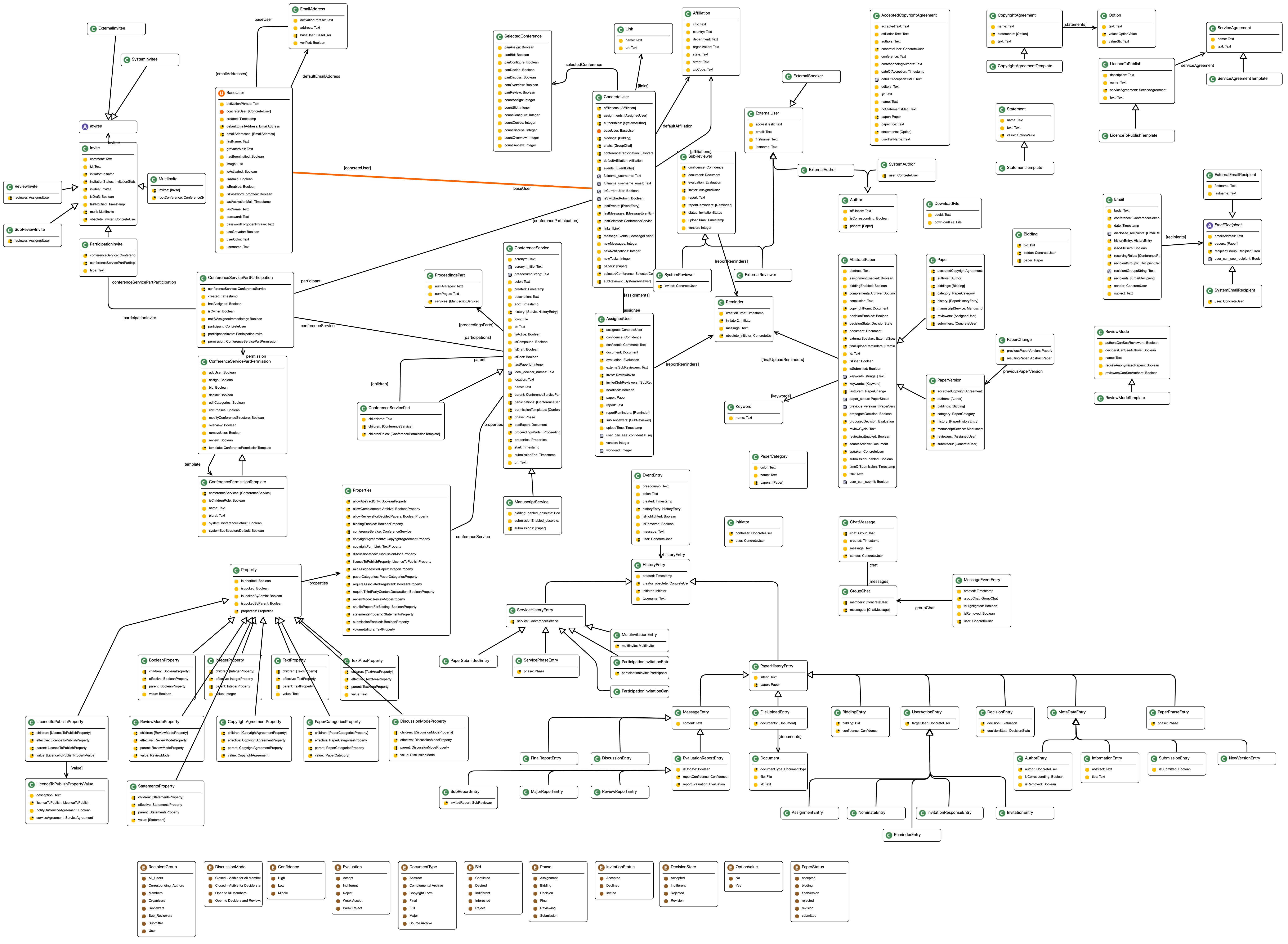}
	\caption{The Data model of Equinocs}
	\label{fig:equinocs-data}
\end{figure}

Thanks to \dime's strict adherence to the \ac{lde} paradigm with its \acp{psl}, despite the number of models involved, each of them clearly illustrates specific aspects of the application, both in terms of the structure of the application's user interface via the \acs{gui} models, and in terms of the application's behavior via the Process models.
As an example, \autoref{fig:equinocs-process-intents} shows a Process model that, based on the entry selected in Equinocs' main menu, directs the control and data flow to the appropriate process model that handles the corresponding page.
\autoref{fig:equinocs-data} shows a bird's eye view of the Equinocs Data model.
While such models tend to grow over time, \dime keeps them performant and accessible. 

Large modeling projects such as Equinocs would be impossible to manage without strong support for hierarchical model structures and consistent integration.
Aligning models with the actual user experience is critical to keeping  application development comprehensible.
Tracing control and data flow in \dime is easy.
Starting from a bird's eye view, the user can iteratively double-click through the submodels to reach basic building blocks such as low-level services.
In addition, \dime{}'s validation views help to monitor the modeling state.  
Due to the complete interconnectedness of the models, any change to a particular model can cause inconsistencies throughout the workspace.
The \emph{Project Validation View} helps identify erroneous models, while the \emph{Model Validation View} points out the specific locations in the model that need to be fixed.
Any break in the control or data flow, as well as inconsistencies in the model hierarchy, are immediately detected and made visible.

\subsubsection{Service Integration}
\label{sec:equinocs-service-integration}

In terms of the application behavior, service integration via \dime{}'s native \acp{sib} allows for the use of existing code libraries and the reuse of complex services, such as the assignment algorithm that computes an appropriate paper-reviewer assignment based on the reviewer's preferences as reflected in the bid.
Another example is the creation of low-level services to send specific \ac{sql} queries at points in the application logic where custom search mechanisms are required.
Regardless of whether small low-level services or entire code libraries have been integrated, at the modeling level the handling of the corresponding native \acp{sib} is seamlessly assimilated into the process modeling workflow as they are just another set of \acp{sib}.

\begin{figure}[tb]
	\centering
	\includegraphics[width=\textwidth]{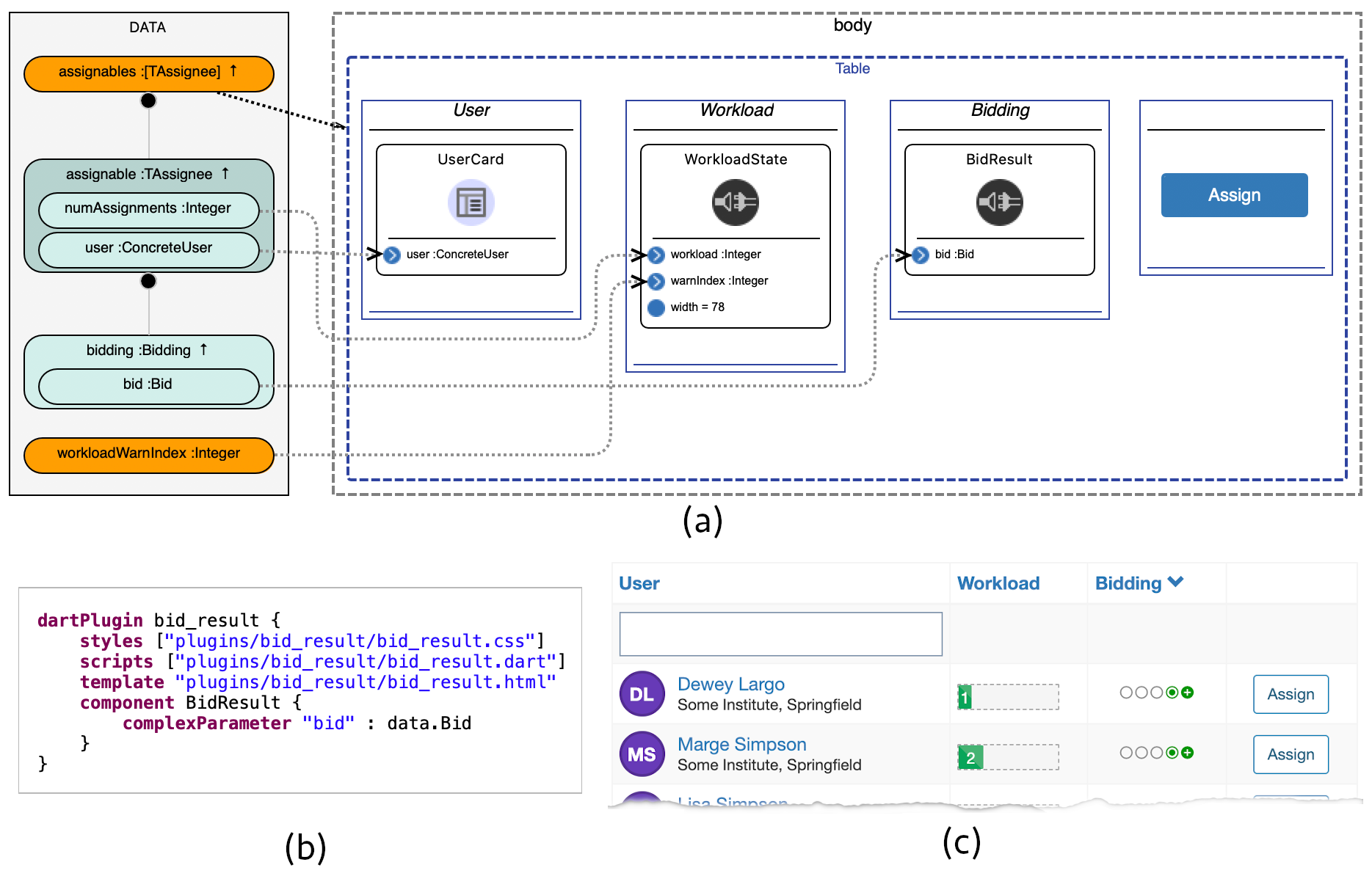}
	\caption{The \textsf{BidResult} \acs{gui} plugin in Equinocs: (a) Usage in a \acs{gui} model, (b) definition in the \acs{gui} Plugin Language, and (c) visualization in the Equinocs system}
	\label{fig:equinocs-plugin}
\end{figure}

In terms of \acs{ui} and \acs{ux}, Equinocs developers took advantage of various model-level customization capabilities for \acs{gui} model components.
In addition, service integration via \dime{}'s \acs{gui} plugins allows for individual customer requirements to be addressed, particularly with regard to the look and feel of \ac{ui} components.
As an example, \autoref{fig:equinocs-plugin} shows the \textsf{BidResult} \acs{gui} plugin, which implements a custom component for visualizing a user's bid for a specific paper.

\subsubsection{Application Deployment}
\label{sec:equinocs-application-deployment}

The deployment of Equinocs to a target environment is based on the structured release management described in \autoref{subsec:dime-app-deploy}.
The automated \ac{ci/cd} pipeline with its resulting Docker images makes it easy to quickly deploy specific Equinocs versions to multiple deployment environments for different purposes.



\section{Conclusion}
\label{sec:conclusion}
We have presented \ac{lde} as a low-code paradigm. 
In particular, we have shown how purpose-specific, graphical modeling enables application experts to design and then produce complex applications, and how an appropriately organized set of \acp{ime} supports collaboration among stakeholders. 
People with different backgrounds contribute to the \emph{one thing} using the \ac{psl} of their mindset, and the model incrementally converges to the intended overall system. 
This elevates the low-code approach to a powerful paradigm for building large systems in a collaborative way. 

We have concretized our approach on the basis of a set of fully automatically generated systems. 
Considering \autoref{fig:overview-ontology}, not only the products (sixth row) but also the \acp{ime} (fourth row) used to build them are fully automatically generated from models.

In fact, their entire deployment was also fully automated based on Rig, our DevOps-supporting \ac{ime}. 
Almost all of our projects are open source, so readers can consult our Website\footnote{\url{https://scce.info}} and our repositories on GitLab\footnote{\url{https://gitlab.com/scce}} to get a more concrete impression of the technology, which rests on two distinctive conceptual pillars: \emph{purpose-specifity} and \emph{full code generation}. 

\ac{lde}'s positioning of \acp{psl} as first-class citizens of system development establishes a new level of reuse, refinement, and evolution. 
Not only can the model evolve over time, but so can the underlying \acp{ime}. 
This results in a continuous improvement cycle where the \acp{ime} become easier to use, which is the ultimate goal of low-code development. 
We are currently working on simplifying this evolution of \acp{ime}, for example by introducing metamodel inheritance and developing a system migration \ac{ime} that aims to (semi-)automatically adapt existing models to changes in the underlying metamodel.
A major challenge in this context is the development and evolution of code generators for the different \acp{ime}.
Therefore, we are developing a dedicated \ac{ime} for this purpose, which combines the service-oriented approach of our earlier work~\cite{JoMaSt2008,Jorges2013} with our recent rule-based transformation technology, similar to Plotkin's Structured Operational Semantics~\cite{KoLyNS2021}.

There are many other directions for future research. 
With the growing abundance of services and \acp{psl}, adequate discovery support becomes important. 
Currently, we use a taxonomy-based classification and retrieval structure for this purpose. 
A more general approach would be to apply recommender systems~\cite{10.1145/3417990.3420200}.

With its new dimension of system development and evolution, \ac{lde} is an exciting area of research with much potential yet to be explored and with enormous practical implications for participation.
The nature of collaborative projects involving stakeholders with different expertises will change radically with \ac{lde}.
The current typically verbal exchange between stakeholders will change to an \ac{ime}-based process of \emph{one thing} construction. 
The required knowledge alignment must be ensured by an adequate constraint-based technology, such as the constraint-based variability modeling framework proposed in~\cite{JoeLMSS2012}.
Making this work in realistic environments requires the combination of many fields, 
such as program analysis and verification, constraint-based synthesis, metamodeling, code generation, test- and 
learning-based validation, software product lines, system evolution, runtime verification, and more.

\bibliographystyle{ACM-Reference-Format}
{
	\bibliography{
		merged-bib}
}

\end{document}